\documentclass[11pt]{article}

\usepackage{verbatim}
\usepackage{amsmath,amscd}
\usepackage{amsfonts}
\usepackage{amssymb}
\usepackage{bbm}
\usepackage{latexsym}
\usepackage{graphicx}
\usepackage{lscape} 
\usepackage{epsfig}
\usepackage{color}
\usepackage{tikz} 
\usepackage{multirow}
\usepackage{afterpage}
\usepackage{amscd}
\usepackage{cite}
\usepackage{tabu} 

\usepackage{hyperref}
\usepackage{longtable}            
\usepackage[font=normalsize]{caption}

\renewcommand{\d}{\mathrm{d}}

\DeclareMathSymbol{\mg}{\mathrel}{symbols}{"1D}

\newcommand{\ga}{\alpha}
\newcommand{\gb}{\beta}
\renewcommand{\gg}{\gamma}
\newcommand{\gd}{\delta}
\renewcommand{\ge}{\epsilon}

\newcommand{\gf}{\phi}
\newcommand{\gvf}{\varphi}

\newcommand{\gm}{\mu}
\newcommand{\gn}{\nu}

\newcommand{\gl}{\lambda}
\newcommand{\gr}{\rho}
\newcommand{\gth}{\theta}

\newcommand{\gs}{\sigma}
\newcommand{\gt}{\tau}
\newcommand{\go}{\omega}

\newcommand{\gp}{\pi}
\newcommand{\gps}{\psi}
\newcommand{\get}{\eta}
\newcommand{\gch}{\chi}
\newcommand{\gD}{\Delta}
\newcommand{\gF}{\Phi}

\newcommand{\gL}{\Lambda}

\newcommand{\gP}{\Pi}
\newcommand{\gPs}{\Psi}

\newcommand{\hI}{\widehat{I}}

\newcommand{\hM}{\widehat{M}}

\newcommand{\hX}{\widehat{X}}

\newcommand{\hget}{\widehat{\eta}}

\newcommand{\hgX}{\widehat{\Xi}}

\newcommand{\hgP}{\widehat{\Pi}}

\newcommand{\cA}{{\cal A}}
\newcommand{\cB}{{\cal B}}
\newcommand{\cC}{{\cal C}}
\newcommand{\cD}{{\cal D}}
\newcommand{\cE}{{\cal E}}
\newcommand{\cF}{{\cal F}}

\newcommand{\cH}{{\cal H}}

\newcommand{\cK}{{\cal K}}

\newcommand{\cM}{{\cal M}}
\newcommand{\cN}{{\cal N}}

\newcommand{\cP}{{\cal P}}

\newcommand{\cS}{{\cal S}}
\newcommand{\cT}{{\cal T}}
\newcommand{\cU}{{\cal U}}
\newcommand{\cV}{{\cal V}}

\newcommand{\cX}{{\cal X}}

\newcommand{\hcE}{\widehat{\cal E}}

\newcommand{\hcH}{\widehat{\cal H}}

\newcommand{\hcM}{\widehat{\cal M}}

\newcommand{\hcP}{\widehat{\cal P}}

\newcommand{\hcR}{\widehat{\cal R}}

\newcommand{\hcT}{\widehat{\cal T}}

\newcommand{\hcZ}{\widehat{\cal Z}}
\newcommand{\tm}{{\tilde m}}
\newcommand{\tn}{{\tilde n}}

\newcommand{\tW}{{\widetilde W}}
\newcommand{\tX}{{\widetilde X}}

\newcommand{\tr}{\text{tr}}

\newcommand{\Id}{\mathbbm{1}}

\newcommand{\ra}{\rightarrow}

\newcommand{\der}{\partial}

\newcommand{\dsp}{\displaystyle}

\newcommand{\beq}{\begin{equation}}
\newcommand{\eeq}{\end{equation}}
\newcommand{\barr}{\begin{array}}
\newcommand{\earr}{\end{array}}
\newcommand{\equ}[1]{\begin{gather} #1 \end{gather}}
\newcommand{\equa}[1]{\begin{align} #1 \end{align}}

\newcommand{\enums}[1]{\begin{enumerate} #1 \end{enumerate}}

\newcommand{\mtrx}[1]{\begin{matrix} #1 \end{matrix}}
\newcommand{\pmtrx}[1]{\begin{pmatrix} #1 \end{pmatrix}}

\newcommand{\non}{\nonumber}
\newcommand{\sfrac}[2]{\mbox{$\frac{#1}{#2}$}}
\newcounter{oldcounter}

\newcommand{\bder}{\bar\partial}
\newcommand{\fM}{\mathfrak{ M}}

\newcommand{\bp}{{\bar p}}
\newcommand{\bq}{{\bar q}}

\newcommand{\byy}{{\bar y}}
\newcommand{\bz}{{\bar z}}

\newcommand{\bW}{{\overline W}}

\newcommand{\bgvf}{{\bar\varphi}}

\newcommand{\bgth}{{\bar\theta}}
\newcommand{\bgs}{{\bar\sigma}}
\newcommand{\bgt}{{\bar\tau}}

\newcommand{\bget}{{\bar\eta}}

\newcommand{\bcA}{{\overline{\cal A}}}

\newcommand{\bcD}{{\overline{\cal D}}}

\newcommand{\Bga}{{\boldsymbol \alpha}}

\newcommand{\Natr}{\mathbbm{N}}
\newcommand{\Intr}{\mathbbm{Z}}
\newcommand{\Cplx}{\mathbbm{C}}
\newcommand{\Real}{\mathbbm{R}}
\newcommand{\Ratl}{\mathbbm{Q}}

\newcommand{\qand}{\quad \text{and} \quad}

\newcommand{\ba}[2]{\[\begin{array}{#2}\label{#1}}
\newcommand{\ea}{\end{array}\]}
\newcommand{\be}{\begin{equation}}
\newcommand{\ee}{\end{equation}}
\newcommand{\bea}{\begin{eqnarray}}
\newcommand{\eea}{\end{eqnarray}}

\newcommand{\E}[1]{\mathrm{E_{#1}}}

\newcommand{\brkt}[2]{\bigl[ ^{#1}_{#2} \bigr]}

\newcommand{\sm}{{\,\mbox{-}}}

\renewcommand{\theequation}{\thesection.\arabic{equation}}

\begin{document}

\thispagestyle{empty}

\begin{flushright}
\end{flushright}
\vskip .2 cm
\begin{center}
{\Large {\bf A Worldsheet Perspective on Heterotic T--Duality Orbifolds} 
}
\\[0pt]

\bigskip
\bigskip {\large
{\bf Stefan Groot Nibbelink}\footnote{
E-mail: s.groot.nibbelink@hr.nl},
\bigskip }\\[0pt]
\vspace{0.23cm}
{\it Institute of Engineering and Applied Sciences, Rotterdam University of Applied Sciences, \\ 
G.J.\ de Jonghweg 4 - 6, 3015 GG Rotterdam, the Netherlands}
\\[2ex] 
{\it Research Centre Innovations in Care, Rotterdam University of Applied Sciences, \\ 
Postbus 25035, 3001 HA Rotterdam, the Netherlands}
\bigskip
\end{center}

\subsection*{\centering Abstract}

Asymmetric heterotic orbifolds are discussed from the worldsheet perspective. 
Starting from Buscher's gauging of a theory of $D$ compact bosons the duality covariant description of Tseytlin is obtained after a non--Lorentz invariant gauge fixing. 
A left--over of the gauge symmetry can be used to removed the doubled constant zero modes so that $D$ physical target space coordinate remain. 
This can be thought of as the worldsheet realization of the strong constraint of double field theory. 
The extension of this description to the heterotic theory is straightforward as all results are written in terms of the invariant and the generalized metrics. 
An explicit method is outline how to obtain a generalized metric which is invariant under T--duality orbifold actions. 
It is explicitly shown how shift orbifolds lead to redefinitions of the Narain moduli. 
Finally, a number of higher dimensional T--folds are constructed including a novel asymmetric $\Intr_6$ orbifold.

\newpage 
\setcounter{page}{1}
\setcounter{footnote}{0}

\tableofcontents 


\section{Introduction and conclusions}
\label{sec:Intro}

\subsection*{Motivation}


The heterotic string~\cite{Gross:1984dd,Gross:1985fr,Gross:1985rr} provides an unified framework to describe all interactions among elementary particles and gravity. Compactifications of this theory to the four dimensions we experience today may be performed on six--dimensional symmetric toroidal orbifolds~\cite{Dixon:1985jw, Dixon:1986jc} without giving up full string computability. Unfortunately, such orbifolds and their related Calabi--Yau compactifications leave many (geometric) moduli unfixed. This impairs predictability of the resulting models as many physical parameters will be ultimately related to undetermined values of these moduli.


Untwisted moduli of such compactifications may be removed by considering asymmetric orbifold actions treating left-- and right--moving string coordinate fields differently by modding out an intrinsically stringy duality symmetry~\cite{Mueller:1986yr}. In the most elementary realization it occurs in a compactification on a circle where the theory at a certain radius $R$ is identified with another compactification on a circle with radius $1/R$ (in string units). In a single construction this is only possible if the radius itself is fixed at the string scale $R=1$. More in general $T$--duality orbifolds result in quotient spaces often referred to as asymmetric orbifolds~\cite{Narain:1986qm}. As such they can be considered as fully computable non--geometric 
string backgrounds~\cite{Hellerman:2002ax,Dabholkar:2002sy,Shelton:2005cf} or so--called $T$--folds~\cite{Hull:2004in,Hull:2006va}. Asymmetric orbifolds have been considered by various research groups~\cite{Harvey:1987da,Ibanez:1987pj,Taylor:1987uv,Narain:1990mw,Imamura:1992np,Imamura:1992bz,Sasada:1994yf,Sasada:1994iv,Erler:1996zs,Aoki:2004sm}; more recent works are e.g.~\cite{Tan:2015nja,Satoh:2016izo,Satoh:2015nlc,Sugawara:2016lpa}.
Asymmetric twists in free fermionic models to stabilize untwisted moduli were exploited in~\cite{Faraggi:2005ib}. The present work focusses on the heterotic setting only, possible dualities to type--II non--geometric compactifications were considered in~\cite{Gautier:2019qiq}. 


In a recent paper~\cite{GrootNibbelink:2017usl} a comprehensive framework to study asymmetric toroidal compactifications of the heterotic string was provided. That work mainly focussed on their construction as generalizations of Narain toroidal compactification modded out by T--duality group elements. The current paper is complementary to that one in the sense, that it aims to  give an explicit duality covariant bosonic worldsheet description of Narain orbifolds. Asymmetric compactifications of the heterotic string are normally discussed on the level of the torus partition function only, without specifying the underlying bosonic worldsheet theory. An explicit worldsheet description is provided by the free--fermionic formulation of the heterotic string~\cite{Antoniadis:1986rn,Antoniadis:1987wp,Kawai:1986ah}. This description predominantly accommodates $\Intr_2$ duality symmetries. 
In order to obtain an explicit bosonic worldsheet description a Buscher's gauging is used and subsequently gauged fixed in such a way that a duality covariant description by Tseytlin is obtained at the expense of manifest worldsheet Lorentz invariance. The basic structure is first exposed for a worldsheet theory of $D$ bosonic fields and after that extended to the heterotic theory with $D$ right--moving and $D+16$ left--moving degrees of freedom.

In addition, this paper gives an extensive description of shift heterotic orbifolds and show that they all can be viewed as Narain torus compactifications. A computational method is provided how the new moduli can be determined from the original ones and the applied shift actions. It is demonstrated in various concrete examples that the orbifold shifts may be geometric or non--geometric and may be accompanied by actions on the gauge part of the Narain lattice.

Finally, since the number of explicitly known truly stringy higher dimensional T--folds is very limited, for examples see e.g.~\cite{Lust:2010iy,Condeescu:2012sp}, this paper provides a number of non--trivial higher dimensional T--fold examples illustrating the developed methodology.

\subsection*{Outline of the paper's main results}

The main results presented in this paper have been structured as follows: 

Section \ref{sc:DualityCovariantWorldsheet} starts from the textbook worldsheet action of $D$ compact bosons. After a Buscher's gauging is applied, a gauge fixing is chosen that breaks manifest Lorentz invariance in favor of achieving a duality covariant doubled worldsheet theory first considered by Tseytlin. A residual gauge symmetry is uncovered that removes the doubling of the constant zero modes. The remaining constant zero modes then have the interpretation of the physical target space coordinates. (This may be thought of as a worldsheet realization of the strong constraint of double field theory~\cite{Hull:2009mi,Hohm:2010jy,Aldazabal:2013sca}.) The worldsheet supersymmetry transformations for this doubled worldsheet theory are derived. In addition, the one--loop partition function is obtained for the doubled theory. Here the boundary terms, which provided the equivalence between the original theory of $D$ compact bosons and Tseytlin's formulation, lead to an additional phase factor that cancels out any dependence on the ``doubled winding numbers''. This ensures that the partition function of the doubled theory is identical to that of the original theory of $D$ compact bosons.

Section \ref{sc:HeteroticExtension} discusses the generalization of these results to the heterotic theory. This generalization is straightforward since all results in Section \ref{sc:DualityCovariantWorldsheet} have been written in terms of the generalized metric and the duality invariant metric which have natural extensions in the heterotic context. On the level of the partition function this reproduces the known results for Narain compactifications of the heterotic string.

Section \ref{sc:NarainOrbifolds} develops the description of orbifold twisting of the Narain theory from the worldsheet point of view. This section recalls the Narain space group description introduced in~\cite{GrootNibbelink:2017usl} and provides a novel way to explicitly construct invariant $\Intr_2$--gradings and associated generalized metrics of Narain orbifolds. If the action is asymmetric, i.e.\ not isomorphic between the right-- and left--movers, so called T--folds are obtained. From the boundary conditions of the worldsheet coordinate fields the full orbifold partition function can be computed. 
One complication is that it is not a priori clear that an invariant generalized metric exists for a given finite orbifold action on the Narain lattice. If not, no asymmetric orbifold can be associated to this action. Here an explicit procedure is outlined how such an invariant generalized metric can be obtained.

Section \ref{sc:ShiftOrbifolds} considers special orbifold actions that have trivial twist parts. It is shown that orbifold shift actions do not lead to new geometries but rather modify the moduli of the Narain compactification. In particular, an explicit description is provided how the new moduli can be computed from the old ones. This is illustrated for a number of simple yet interesting cases such as geometrical and non--geometrical shifts combined with non--trivial Wilson lines.

Finally, Section \ref{sc:T-folds} provides a number of examples of higher dimensional T--folds. The first two examples are Narain orbifolds obtained by applying the basic T--duality twist to all $D$ compact dimensions. One acts only on the right--movers hence it fixes all moduli but is only supersymmetric in $D=4$ or $8$ dimensions, while the other only acts on $D$ left--movers preserving all supersymmetries in any dimension and the Wilson lines are left free. In a further example a $\Intr_6$ action is realized in an asymmetric way. This provides one of the first explicitly known examples of higher order asymmetric orbifolds in four dimensions.

This paper is concluded with three Appendices that provide some technical background for the results obtained in this work. Appendix~\ref{sc:EuclideanPathintegrals} introduces the notation used to evaluate partition functions. Appendix~\ref{sc:ModFunTrans} describes the underlying modular transformations. Finally, Appendix~\ref{sc:Narain Geometry} gives further details of the description of Narain moduli and their transformations as uncovered in~\cite{GrootNibbelink:2017usl}.

\subsection*{Acknowledgements}

The author would like to thank  P.K.S.\ Vaudrevange for many enlightening discussion on Narain orbifolds which provided the starting point for this work. The author would also like to thank O.\ Loukas for carefully reading the manuscript. \\
In addition, the author would like to thank H.P.\ Nilles and S.\ Ramos-S\'anchez for enlightening discussions and pointing out some minor issues in the published version of the manuscript.

\section{Duality Covariant Worldsheet}
\label{sc:DualityCovariantWorldsheet} 
\setcounter{equation}{0}


\subsection{Doubled Target Space Torus}

The Minkowskian worldsheet is parameterized by the worldsheet time $\gs_0$ and space $\gs_1$ coordinates. The starting point of the description of $D$ bosons $X^T = \big(X^1, \ldots, X^D)$, the internal coordinates fields, on the worldsheet is the action 
\equ{ \label{eq:DbosonAction} 
S = 
\int \dfrac{\d^2\gs}{2\pi}
\Big[ 
\dfrac 12\, \der_0 X^TG\, \der_0 X 
- \dfrac 12\, \der_1 X^T G\, \der_1 X 
+ \der_1 X^T B\, \der_0 X
\Big]~, 
}
%
where $G$ is a $D$--dimensional metric of and $B$ an anti--symmetric tensor on a target space torus $T^D$. Throughout this work these background quantities are taken to be constant. The target space torus periodicities are encoded in integral lattice identifications  
\equ{ \label{eq:DTorusPeriodicities}
X \sim X+ 2\pi\, m~, 
\quad 
m \in \Intr^D~; 
}
the geometrical aspects of the torus $T^D$ have already been taken into account by the metric $G$ in the worldsheet action~\eqref{eq:DbosonAction}. 

A duality covariant description of this theory can be obtained following Buscher's gauging procedure~\cite{Buscher:1987sk} and subsequently choosing an appropriate gauge~\cite{Rocek:1997hi}: The coordinate fields $X$ are promoted to possess the following gauge transformations 
\equ{ \label{eq:GaugeTransX}
X \mapsto X - \gl~, 
}
where $\gl^T= \big(\gl^1, \ldots, \gl^D)$ are general functions on the worldsheet. To ensure invariance of the action~\eqref{eq:DbosonAction} the derivatives are promoted to gauge covariant ones 
\equ{ 
\cD_\gm X = \der_\gm X + \cA_\gm~,  
}
(where $\gm=0,1$) which are gauge invariant, provided that the gauge fields $\cA_\gm$ themselves transform as 
\equ{ \label{eq:GaugeTransA}
\cA_\gm \mapsto \cA_\gm + \der_\gm \gl~. 
}
This gauging would remove all physical bosons from the worldsheet. To avoid this, the gauged action is complemented by a Lagrange multiplier field $\tX$\,, which enforces that the gauge field is pure gauge: 
\equ{ \label{eq:GaugedAction} 
S = 
\int \dfrac{\d^2\gs}{2\pi}
\Big[ 
\dfrac 12\, \cD_0 X^T\,G \cD_0 X 
- \dfrac 12\, \cD_1 X^T G\, \cD_1 X 
+ \cD_1 X^T B\, \cD_0 X 
+ \tX^T \cF_{10}
\Big]~,  
}
where $\cF_{\gm\gn} = \der_\gm \cA_\gn-\der_\gn\cA_\gm$ is the gauge field strength. The Lagrange multiplier fields $\tX$ satisfy similar periodicities as the coordinates $X$ themselves: 
\equ{ \label{eq:DualTorusPeriodicities}
\tX \sim \tX + 2\pi\, \tm~, 
\quad 
\tm \in \Intr^D~,
}
so that the charges 
\equ{ \label{eq:Quantization} 
q = \int \dfrac{\cF_{2}}{2\pi} \in \Intr^D
}
are integral in the Euclidean theory, hence the periodicities~\eqref{eq:DualTorusPeriodicities} of $\tX$ result in a trivial phase $\exp\{-2\pi i\, \tm^Tq\}=1$ in the path integral. A gauge, that makes the duality manifest, is~\cite{Rocek:1997hi}:\footnote{Another gauge choice would be $\cA_0=0$; but for the current purposes this choice would be less convenient.} 
\equ{ \label{eq:GF_A1=0} 
 \cA_1(\gs_1,\gs_0) \stackrel{!}{=} 0~.
}
Clearly, this gauge breaks manifest Lorentz invariance on the worldsheet. In this gauge $\cD_1 X = \der_1 X$\,, $\cD_0 X = \der_0 X + \cA_0$ and $\cF_{10} = \der_1 \cA_0$. Inserting this in the action~\eqref{eq:GaugedAction} and performing a partial integration to remove the derivative $\der_1$ on the remaining gauge field component $\cA_0$, shows that the equation of motion of $\cA_0$ is algebraic: 
\equ{ 
\der_0 X + \cA_0 = G^{-1}\big( \der_1 \tX + B\,\der_1 X\big)~. 
}
Eliminating all $\cA_0$ dependence using this expression, shows that the action can now be cast in the Tseytlin's form~\cite{Tseytlin:1990nb,Tseytlin:1990va} by two further partial integrations: 
\equ{ \label{eq:DualityCovariantAction} 
S= \int \dfrac{\d^2\gs}{2\pi}
\Big[ 
- \dfrac 12\, \der_1 Y^T\, \hcH\, \der_1Y + \dfrac 12\, \der_1 Y^T\, \hget\, \der_0 Y
\Big]~, 
}
where $Y^T = \big(X^T ~ \tX^T\big)$ combines the coordinates $X$ and the dual coordinates $\tX$ in a single $2D$-dimensional vector. In addition, the generalized metric $\hcH$ and the $\text{O}(D,D;\Real)$ invariant metric $\hget$ are introduced 
\equ{ \label{eq:GeneralizedODDMetrics}
\hcH = 
\pmtrx{
G - BG^{-1}B & -BG^{-1} \\[1ex] 
G^{-1} B & G^{-1} 
}
\qand
\hget = 
\pmtrx{ 
0 & \Id_D \\[1ex] 
\Id_D & 0 
}~. 
}
Given the periodicities, \eqref{eq:DTorusPeriodicities} and~\eqref{eq:DualTorusPeriodicities} of the coordinates fields $X$ and their duals $\tX$, the doubled coordinates $Y$ are subject to the periodicities 
\equ{ 
Y \sim Y + 2\pi\, N~, 
\quad 
N = \pmtrx{ m \\ \tm} \in \Intr^{2D}~. 
}
Just as the periodicities of the dual coordinates $\tX$ were enforced by charge quantization~\eqref{eq:Quantization}, the periodicities of $X$ can be understood in the same fashion, as by a duality transformation the roles of the coordinates $X$ and their duals $\tX$ can be interchanged. 
Consequently, the Tseytlin action~\eqref{eq:DualityCovariantAction} is invariant under the $\hM \in \text{O}_{\hget}(D,D; \Intr)$ duality transformations
\equ{ 
Y \mapsto \hM^{-1}\, Y~, 
\qquad 
\hcH \mapsto \hM^T\, \hcH\, \hM~, 
}
since by definition $\hM^T \,\hget\, \hM = \hget$ and the lattice $\Intr^{2D}$ is mapped to itself: $\hM^{-1}\, \Intr^{2D} = \Intr^{2D}$\,.

In addition, the generalized metric and the $\text{O}(D,D;\Real)$--invariant metric~\eqref{eq:GeneralizedODDMetrics} satisfy the following properties 
\equ{ \label{eq:PropertiesGeneralizedMetric} 
\hcH^T = \hcH~, 
\quad 
\hcH\, \hget^{-1}\, \hcH = \hget~. 
}
This allows to define a $\Intr_2$--grading
\equ{ \label{eq:Z2Grading} 
\hcZ = \hget^{-1} \hcH = \hcH^{-1} \hget~, 
\quad 
\hcZ^2 = \Id_{2D}~,
\quad 
\hcZ^T \, \hget\, \hcZ\, = \hget~, 
\quad
\tr_{2D}\big[ \hcZ \big] = 0~. 
}
The one but last relation implies that $\hcZ$ itself is an element of the duality group with real coefficients: $\hcZ \in \text{O}(D,D;\Real)$.

In the derivation of~\eqref{eq:DualityCovariantAction} three partial integrations were performed. Since the coordinate fields $X$ and duals $\tX$ are quasi--periodic (but not periodic) in general, the resulting boundary terms 
\equ{
S_\text{bnd} = \int \dfrac{\d^2\gs}{2\gp}
\Big[
\der_1\big(\tX^T\cA_0\big) + \dfrac 12\,\der_1\big(\tX^T\der_0 X\big) - \dfrac 12\, \der_0\big(\tX^T\der_1 X\big)
\Big]~.
}
do not automatically vanish. In particular, because of~\eqref{eq:DualTorusPeriodicities} the first term gives a boundary contribution
\equ{ 
\int \dfrac{\d^2\gs}{2\gp}\, \der_1\big(\tX^T\cA_0\big) 
\sim \int \d\gs_0\, \tm^T\cA_0(0, \gs_0) 
\stackrel{!}{=}0~,
}
which can be set to zero by a further gauge fixing. Indeed, the gauge fixing~\eqref{eq:GF_A1=0} does not fix the gauge completely; there are residual gauge transformations with gauge parameters  $\gl=\gl(\gs_0)$\,, which are functions of the worldsheet time $\gs_0$ only. Using this residual gauge transformation, a further gauge fixing 
\equ{ \label{eq:GF_A00=0}
\cA_0(0,\gs_0) \stackrel{!}{=} 0
}
can be enforced. In the combined gauge~\eqref{eq:GF_A1=0} and~\eqref{eq:GF_A00=0} the boundary action reduces to 
\equ{ \label{eq:BoundaryAction} 
S_\text{bnd} = \int \dfrac{\d^2\gs}{2\gp}
\Big[
 \dfrac 12\,\der_1\big(\tX^T\der_0 X\big) - \dfrac 12\, \der_0\big(\tX^T\der_1 X\big)
\Big]~.
}

Since the gauge transformation~\eqref{eq:GaugeTransA} of the gauge fields involves derivatives, even this does not fix the gauge completely: Constant shifts $\gl=\gl_0$ in~\eqref{eq:GaugeTransX} are still allowed. This means that the doubled coordinates $Y$ used in~\eqref{eq:DualityCovariantAction} are uniquely defined up to constant shifts in $D$ directions 
\equ{ \label{eq:DoubledCoordConstantShifts}
Y \sim Y - \gL_0~,
\quad 
\gL_0 = \hM^{-1}\, \pmtrx{ \gl_0 \\ 0 }~,
}
for some $\hM\in\text{O}_{\hget}(D,D;\Intr)$ defining the used duality frame and $\gl_0 \in \Real^D$. In other words of the $2D$ constant zero--modes of the doubled coordinate fields $Y$ only $D$ are physical, assuming that~\eqref{eq:DbosonAction} should be taken as the starting point of the worldsheet description.

\subsection{Worldsheet Supersymmetry}
\label{sc:ODDSusyWS} 

The worldsheet action~\eqref{eq:DbosonAction} can be extended to 
\equ{ \label{eq:SusyAction} 
S = 
\int \dfrac{\d^2\gs}{2\pi}
\Big[ 
- \der X^T K\, \bder X 
- \gps^T\, \der \gps
\Big]~,
\quad\text{where}\quad K = G+B~,  
}
with $D$ right--moving real fermions $\gps$\,. Here left-- and right--moving coordinates $(\gs,\bgs)$ and their associated derivatives $(\der,\bder)$,  
\begin{subequations}
\equa{ 
\gs= \dfrac1{\sqrt 2} \big( \gs_1+\gs_0\big) 
&\qand
\bgs = \dfrac1{\sqrt 2} \big(\gs_1- \gs_0\big)~, 
\\[1ex] 
\der = \dfrac 1{\sqrt 2} \big(\der_1 + \der_0 \big) 
&\qand
\bder = \dfrac 1{\sqrt 2} \big( \der_1 - \der_0\big)~, 
}
\end{subequations} 
were introduced, so that the two--dimensional measure can be written as
\(
\d^2 \gs = \d \gs_1 \d \gs_0 = \d \bgs \d\gs\,.  
\)
This action is invariant under the supersymmetry transformations 
\equ{ \label{eq:RightSusy}
\gd X = \ge\, e^{-1}\, \gps
\qand
\gd \gps = - \ge\, e\, \bder X~, 
}
where $e$ is a vielbein associated the metric $G = e^Te$\,.

Applying the Buscher's gauging to this action leads to 
\equ{ 
S = 
\int \dfrac{\d^2\gs}{2\pi}
\Big[ 
- \cD X^T K\, \bcD X 
- \gps^T\, \der \gps 
+ \tX^T \cF 
\Big]~, 
}
where $\cD X = \der X + \cA$\,, $\bcD X = \bder X + \bcA$ and $\cF = \cF_{10} = \bder \cA - \der \bcA$. This action is still supersymmetric, provided that the transformations~\eqref{eq:RightSusy} are extended to 
\equ{ 
\gd X = \ge\, e^{-1}\, \gps~, 
\quad
\gd \gps = - \ge\, e\, \bcD X
\qand 
\gd \tX = - \ge\, \big( e^T + B e^{-1} \big)\, \gps~. 
}
Since the Tseytlin's form of the action was obtained by using the gauge $\cA_1=0$, the supersymmetry variation of $\gps$ then becomes 
\equ{ 
\gd \gps = - \dfrac1{\sqrt 2}\, \ge\, 
\Big[ \big(e - e^{-T} B\big)\,\der_1 X - e^{-T} \, \der_1\tX \Big]~. 
}
Hence, using the doubled coordinate $Y$, the supersymmetry transformations can be cast in the form: 
\equ{ \label{eq:DoubledRightSusy}
\gd Y = \ge 
\pmtrx{ e^{-1} \\[1ex] -e^T - B e^{-1} } \gps
\qand 
\gd \gps = - \dfrac 1{\sqrt 2} \ge 
\pmtrx{ e-e^{-T} B & - e^{-T} } \der_1 Y~.  
}

These transformations can also be obtained directly from the duality covariant action~\eqref{eq:DualityCovariantAction} extended with the right--moving fermions $\gps$: 
\equ{ \label{eq:DualityCovariantActionLR} 
S= \int \dfrac{\d^2\gs}{2\pi}
\Big[ 
- \dfrac 1{\sqrt 2}\, \der_1 Y^T
\Big(  \dfrac {\hcH+\hget}2\, \bder Y +  \dfrac{\hcH-\hget}2\, \der Y \Big)
- \gps^T \der \gps
\Big]~, 
}
where the left-- and right--moving derivatives $\der$ and $\bder$ of the doubled coordinates $Y$ have been separated. Since the action of the right--moving fermions $\gps$ involve the left--moving derivative $\der$ only,  any supersymmetry that involves these fermions can only be related to the term involving the operator $\frac 12\big(\hcH - \hget)$ since that term also contains this derivative. By introducing the generalized vielbein
\equ{ \label{eq:ODDVielbein}
\cE = \cK\, \hcE~,  
\quad 
\cK = \dfrac 1{\sqrt 2}\,\pmtrx{\Id_D & -\Id_D \\[1ex] \Id_D & \phantom{-}\Id_D}
\qand 
\hcE = \pmtrx{
e & 0 \\[1ex] 
e^{-T} B & e^{-T} 
}~,  
}
satisfying
\equ{
\cK^{-T}\, \hget\, \cK^{-1} = \get =
\pmtrx{-\Id_D & 0 \\[1ex] 0 & \Id_D}~, 
\quad 
\hcE^T\, \hget\, \hcE = \hget
\qand 
\cE^T\, \cE 
= \hcH~, 
}
where $\get$ is the Minkowskian metric with signature $(D,D)$, 
the matrices 
\equ{ \label{eq:SplittingOperators}
\dfrac{\hcH + \hget}2  = \hcH \, \dfrac{\Id+\hcZ}2 = \cE^T\, \cP_\text{L}\, \cE
\qand 
\dfrac{\hcH - \hget}2 = \hcH \, \dfrac{\Id-\hcZ}2 = \cE^T\, \cP_\text{R}\, \cE
} 
can be related to the left-- and right--projections defined by the $\Intr_2$--grading $\hcZ$ leading to the projection operators 
\equ{ \label{eq:LRProjectors} 
\cP_\text{L} = \dfrac{\Id_{2D} + \get}2 = 
\pmtrx{0 & 0 \\[1ex] 0 & \Id_D}
\qand 
\cP_\text{R} = \dfrac{\Id_{2D} - \get}2 = 
\pmtrx{\Id_D & 0 \\[1ex] 0 & 0}~.  
}
Inserting~\eqref{eq:SplittingOperators} in the action~\eqref{eq:DualityCovariantActionLR} leads to 
\equ{ \label{eq:DualityCovariantActionLRSplitted} 
S= \int \dfrac{\d^2\gs}{2\pi}
\Big[ 
- \dfrac 1{\sqrt 2}\, \der_1 Y^T \cE^T
\Big(  \cP_\text{L}\, \cE\, \bder Y + \cP_\text{R}\, \cE\,\der Y \Big)
- \gps^T \der \gps
\Big]~. 
}
This suggests the following supersymmetry transformations
\equ{ 
\gd Y = \ge\, \cE^{-1} \cP_R \cV\, \gps
\qand
\gd \gps = \dfrac{1}{\sqrt 2}\, \ge\, \cU^T \cP_\text{R} \cE\, \der_1 Y~, 
}
in terms of two $2D\times D$--matrices $\cU$ and $\cV$. These transformations leave the action~\eqref{eq:DualityCovariantActionLRSplitted} invariant, provided, that $\cU= - \cV$. Moreover, given the form of the right--moving projector $\cP_\text{R}$, given in~\eqref{eq:LRProjectors}, one may set 
\equ{ 
\cP_\text{R} \cV = \cV = \sqrt{2}\, \pmtrx{ \Id_D \\[1ex] 0}~. 
}
Making this choice, the supersymmetry transformations~\eqref{eq:DoubledRightSusy} are recovered using the expression of the generalized vielbein~\eqref{eq:ODDVielbein}. The closure of the supersymmetry transformations on the fields $\gps$ and $Y$ reads 
\equ{
[\gd_1, \gd_2] \gps = 2\sqrt{2}\, \ge_1\ge_2\, \der_1 \gps
\qand
[\gd_1, \gd_2] Y = 2\sqrt{2}\, \ge_1\ge_2\, \dfrac{\Id - \hcZ}2\, \der_1 Y~. 
}

\subsection{One--Loop Partition Function}

To determine the one--loop partition function by computing the path integral on the worldsheet torus, 
\equ{ \label{eq:ODDPathintegral} 
Z_\text{Doubled} = \int \cD Y \cD h \cD \cA_1 \cD b \cD c\, 
e^{-S_\text{Eucl}}~, 
}
a quantum version of the Tseytlin's action~\eqref{eq:DualityCovariantAction} is required. As this is a gauge fixed action, following the standard BRST--procedure the full quantum Euclidean action reads 
\equ{ \label{eq:EuclWSAction}
S_\text{Eucl} 
= \int \dfrac{\d^2z}{2\pi}
\Big[ 
\dfrac 14\, \der_1 Y^T \hcH\, \der_1 Y + \dfrac i4\, \der_1 Y^T \hget\, \der_2 Y 
+ \dfrac 1{\sqrt 2}\,h^T \cA_1 + \dfrac 1{\sqrt 2}\,b^T\der_1 c
\Big]~, 
}
where $h$ are $D$ Lagrange multipliers enforcing the gauge $\cA_1=0$ and $b, c$ form $D$ associated ghost systems. The fields $h$ and $\cA_1$ can be trivially integrated out without leaving a trace.

The one--loop periodic boundary conditions for the ghosts $b, c$ and the quasi--periodicities 
\equ{ 
Y(z+1) = Y(z) + 2\pi\, N
\qand 
Y(z-\gt) = Y(z) + 2\pi\, N'~, 
}
for the doubled coordinate fields $Y$ with $N, N' \in \Intr^{2D}$\,, are solved by the off--shell mode expansions 
\equ{ 
\mtrx{ \dsp 
b(z) = \left.\sum_{r,r'}\right.^\prime  \gF\brkt{r}{r'}(z)\,b\brkt{r}{r'}~, 
\\[1ex] \dsp 
c(z) = \left.\sum_{r,r'}\right.^\prime  \gF\brkt{r}{r'}(z)\,c\brkt{r}{r'}~, 
}
\qand 
Y(z) = 2\pi\, \gf\brkt{N}{N'}(z) + \left.\sum_{r,r'}\right.^\prime  \gF\brkt{r}{r'}(z)\,Y\brkt{r}{r'}~, 
}
using the definitions~\eqref{eq:LinearMode} and~\eqref{eq:QuasiPeriodicMode}. The prime on the sum denotes the sum over all integers $r,r'\in \Intr$ excluding the zero--mode $r=r'=0$ contribution. Inserting these mode expansions in the remaining path integral and evaluating the infinite dimensional integrals over the mode coefficients $b\brkt{r}{r'}$\,, $c\brkt{r}{r'}$ and $Y\brkt{r}{r'}$ leads to an expression involving infinite products: 
\equ{ \label{eq:PartitionODDintermediate}
Z_\text{Doubled} = 
\left.\prod_{r,r'}\right.^\prime
\left[ 
\dfrac{\det{}_{2D}\big[ \frac {r}{\sqrt 2} \Id_{2D} \big]}
{\det{}_{2D}\big[(r)\big\{(r\,\gt_1 + r')\, \hget + r\, i\gt_2\, \hcH  \big\} \big]}
\right]^{1/2}\!\!\!\!
\sum_{N,N'\in \Intr^{2D}}
(-1)^{N^T\, \hget\, N'}\, 
e^{2\pi i\, \frac 12\, N^T \big( \gt_1\, \hget + i\gt_2\, \hcH \big) N}~. 
} 
Notice that the infinite product factors $\left.\prod\right.^\prime r$ due to the ghosts and the $\der_1$--derivative on $Y$ in the worldsheet action~\eqref{eq:EuclWSAction} cancel. In fact, since this is a pure constant, i.e.\ not $\gt$--dependent, infinite factor, it may be dropped from the path integral altogether.\footnote{The ghost contributions do not dropped so easily, when the Buscher's gauge $\cA_0=0$ had been chosen,  since the resulting infinite product factors would be $\gt$--dependent.}

Using the properties~\eqref{eq:PropertiesGeneralizedMetric} and~\eqref{eq:Z2Grading} it follows, that 
\equ{ \label{eq:DetGeneralizedODDMetrics}
\det{}_{2D} \big[ a\, \hget + b\, \hcH \big] = (-)^D \big(a^2 - b^2\big)^D~, 
}
 is independent of the moduli $G$ and $B$\, for any two complex constants $a, b \in \Cplx$\,. (A derivation of this result in a more general Narain context can be found in appendix~\ref{sc:GenMetricIndependence}.) Consequently, the remaining infinite product factor reduces to 
\equ{ 
\left.\prod_{r,r'}\right.^\prime
\left[ 
\dfrac{1}
{\det{}_{2D}\big[(r\,\gt_1 + r')\, \hget + r\, i\gt_2\, \hcH \big]}
\right]^{1/2}
= 
\left.\prod_{r,r'}\right.^\prime \dfrac{1}
{| r\, \gt + r'|^D}
= \dfrac 1{
|\get(\gt)|^{2D}}~, 
}
which can be expressed in terms of the Dedekind eta--function $\get(\gt)$ using~\eqref{eq:ProductDedekind}.

The phase $(-1)^{N^T\,\hget\,N' }$ in~\eqref{eq:PartitionODDintermediate} is modular invariant by itself. The appearance of this phase may seem somewhat surprising, since computing the partition function for $X$ directly would not lead to this phase factor. The cause of this can be traced back to the observation that the worldsheet actions~\eqref{eq:DbosonAction} and~\eqref{eq:DualityCovariantAction} are equivalent to each other up to the boundary contribution~\eqref{eq:BoundaryAction}, which can be evaluated on the Euclidean worldsheet torus to 
\equ{ \label{eq:EuclBnd}
S_\text{Eucl.\,bnd} = -i\int\dfrac{\d x \d y}{2\pi}\, \Big[
\der_x \big( \tX^T \der_y X\big) 
-\der_y \big( \tX^T \der_x X\big)\Big] = 
-2\pi\, \dfrac i2\, \big( \tm^Tm' - m^T \tm'\big)~,  
}
using the worldsheet torus coordinates $x,y$ defined in~\eqref{eq:StraightWSTcoord}. This thus leads to precisely the same phase in the path integral and hence they cancel out. (In any event, since partition functions are defined from the path integral up to modular invariant phases, we are always free to include this factor once more, so that it cancels out.) Including this phase means that the sum over $N'$ is trivial giving rise to an infinite constant factor, which may subsequently be dropped. 

Observe that this is very different to what happens to the quantum numbers $n'\in\Intr^D$ that label the worldsheet boundary conditions in the $\gt$--direction in the original theory of a $D$--dimensional target space torus. In that case $n'$ are physical, as they can be interpreted as the Poisson resummed Kaluza--Klein numbers. In the doubled formalism both winding and Kaluza--Klein quantum numbers are contained in $N$ simultaneously and hence there is no need for $N'$. 

The partition function can then finally be written as 
\equ{ \label{eq:PartitionODD} 
Z_\text{Doubled} =
\dfrac{1}
{
|\get(\gt)|^{2D}}\, 
\sum_{N \in \Intr^{2D}} 
e^{2\pi i\, \frac 12\, N^T \big( \gt_1\, \hget + i\gt_2\, \hcH\big) N}
~, 
}
It is possible to express this partition function as 
\equ{
Z_\text{Doubled} = \dfrac{1}
{
|\get(\gt)|^{2D}}\sum_{N \in \Intr^{2D}} 
q^{\frac 12 P_L^2} \bq^{\frac 12 P_R^2}
~, 
}
in terms of left-- and right--moving momenta,
\equ{
P_\text{L} = \cP_\text{L} \, P~, 
\quad 
P_\text{R} = \cP_\text{R}\, P
\qand 
P = \cE\, N~,
}
are defined using the left-- and right--moving projectors~\eqref{eq:LRProjectors} and the generalized vielbein~\eqref{eq:ODDVielbein}.

\subsection{Relation to Double Field Theory}

Double field theory~\cite{Hull:2009mi,Hohm:2010jy,Aldazabal:2013sca}  is an attempt to obtain a duality covariant target space description of string theory. There the dimension $D$ of a target space torus is doubled to $2D$ and the generalized metric is assumed to be the metric on the doubled torus. However, since only $D$ coordinates are physical, a so--called {\em strong constraint} is being implemented by hand to remove $D$ of the $2D$ doubled coordinates. 

The doubled worldsheet description of strings on a $D$--dimensional torus presented here should not be confused with a worldsheet theory where the target space is a torus of the dimension $2D$. There are several important differences that appeared by performing the Buscher's gauging procedure and the subsequent gauge fixing: 
\enums{
\item The duality covariant action~\eqref{eq:DualityCovariantAction} is not manifestly Lorentz invariant;
\item The doubled coordinates $Y$ are defined modulo constant shifts~\eqref{eq:DoubledCoordConstantShifts} in $D$ of the $2D$ doubled directions; 
\item Infinite product factors \smash{$\prod\limits_{r,r'} r$} were cancelled by ghost contributions in the path integral~\eqref{eq:PartitionODDintermediate}; 
\item All dependence on the integral vector $N'$ dropped out because of a phase that arose from~\eqref{eq:EuclBnd}. 
}
Hence, in particular, the removal of $D$ of the $2D$ doubled torus coordinates is not enforced by hand, but rather is a left--over consequence of the gauge fixing procedure, which did not fix the gauge completely. The other consequences mentioned here have no interpretation in the target space theory: A (none Lorentz invariant) worldsheet action on an one--loop torus with two different cycles is simply not part of the target space description.

\section{Heterotic Extension}
\label{sc:HeteroticExtension} 
\setcounter{equation}{0}

\subsection{Narain Lattice Worldsheet }

The extension to the heterotic string theory can be obtained by replacing the  $\text{O}_{\hget}(D,D;\Real)$--invariant metric $\hget$ (as given in~\eqref{eq:GeneralizedODDMetrics}) and the generalized metric $\hcH$ by their heterotic counter parts: 
\equ{ \label{eq:ODD16Metric}
\hget = 
\pmtrx{
0 & \Id_D & 0 \\[1ex] 
\Id_D & 0 & 0 \\[1ex]
0 & 0 & g_{16} 
}~, 
}
where the 16--dimensional Cartan metric $g_{16}$ is given by~\eqref{eq:E88metric}, and 
\equ{ \label{eq:GeneralizedNarainMetric}
\hcH = 
\pmtrx{
G + A^T A + C G^{-1} C^T & -C G^{-1} & (\Id_D + C G^{-1}) A^T \ga_{16}
\\[1ex] 
-G^{-1} C^T & G^{-1} & - G^{-1} A^T \ga_{16}
\\[1ex] 
\ga_{16}^T A (\Id_D+ G^{-1} C^T) & -\ga_{16}^T A G^{-1} & \ga_{16}^T \big(\Id_{16}+ A G^{-1} A^T \big) \ga_{16}
}~, 
}
with $C = B + \sfrac 12\, A^T A$\,, in the Tseytlin's action. This leads to 
\equ{ \label{eq:TseylinNarainAction} 
S= \int \dfrac{\d^2\gs}{2\pi}
\Big[ 
- \dfrac 12\, \der_1 Y^T \hcH\, \der_1Y + \dfrac 12\, \der_1 Y^T \hget\, \der_0 Y 
-(\der_0 x)^2 +(\der_1 x)^2 
- \gPs^T \der \gPs
- \gps^T \der \gps
\Big]~, 
}
where the generalized coordinate vector $Y^T = \big(X^T ~ \tX^T ~ \gch^T \big)$ is extended to include 16 bosonic gauge degrees of freedom $\gch$\,, satisfying the torus periodicities 
\equ{ \label{eq:NarainLattice} 
Y \sim Y + 2\pi\, N~, 
\quad 
N^T = \pmtrx{ m^T & \tm^T & q^T} \in \Intr^{2D+16}~, 
}
of a $2D+16$--dimensional Narain lattice. The full heterotic worldsheet theory is completed in light--cone gauge by $d=8-D$ real bosons $x$. Furthermore, $\gps$ and $\gPs$ are the $d$ and $D$ real fermions, that are the super partners of $x$ and $X$, respectively. In addition to the $D$--dimensional metric $G$ and the anti--symmetric tensor $B$, the generalized metric $\hcH$ now also depends on the $16\times D$ component Wilson line matrix $A$, corresponding to 16--component Wilson lines in $D$--directions.

The properties of the $\text{O}(D,D+16;\Real)$--invariant metric $\hget$,  the generalized metric $\hcH$ and the $\Intr_2$--grading $\hcZ = \hget^{-1} \hcH = \hcH^{-1} \hget \in \text{O}(D,D+16;\Real)$  are very similar to those before:
\equ{ \label{eq:Z2GradingProperties} 
\hcH^T = \hcH~, 
\quad 
\hcH\, \hget^{-1}\, \hcH = \hget~, 
\quad 
\quad 
\hcZ^2 = \Id~,
\quad
\tr\big[ \hcZ \big] = 16~, 
}
(where $\Id = \Id_{2D+16}$) except for the last one, which reflects that there is a mismatch in left-- and right--moving bosonic degrees of freedom in the heterotic theory.

\subsection{Worldsheet Supersymmetry}

The action~\eqref{eq:TseylinNarainAction} can be written as
\equ{ \label{eq:NarainActionLR} 
S= \int \dfrac{\d^2\gs}{2\pi}
\Big[ 
- \dfrac 1{\sqrt 2}\, \der_1 Y^T
\Big(  \dfrac {\hcH+\hget}2\, \bder Y +  \dfrac{\hcH-\hget}2\, \der Y \Big)
+ \dfrac 12\, (\der_1 x)^2 - \dfrac 12\,(\der_0 x)^2 
- \gPs^T \der \gPs
- \gps^T \der \gps
\Big]~.
}
Also in the Narain case a generalized vielbein can be defined by~\eqref{eq:NarainVielbein} satisfying~\eqref{eq:MinkowskiDD16Metric}. Hence, the matrices 
\equ{ \label{eq:SplittingNarainOperators}
\dfrac{\hcH + \hget}2 = \hcH \, \hcP_L = \cE^T\, \cP_\text{L}\, \cE
\qand 
\dfrac{\hcH - \hget}2 = \hcH \, \hcP_R = \cE^T\, \cP_\text{R}\, \cE~, 
} 
are related to left-- and right--projection operators 
$\hcP_L = \frac 12(\Id+\hcZ)$ and $\hcP_R = \frac 12(\Id-\hcZ)$
and 
\equ{ \label{eq:LRNarainProjectors} 
\cP_\text{L} = \dfrac{\Id_{2D} + \get}2 = 
\pmtrx{0 & 0 & 0 \\[1ex] 0 & \Id_D & 0  \\[1ex] 0 & 0 & \Id_{16} }
\qand  
\cP_\text{R} = \dfrac{\Id_{2D} - \get}2 = 
\pmtrx{\Id_D & 0 & 0 \\[1ex] 0 & 0 & 0 \\[1ex] 0 & 0 & 0}~, 
}
and can therefore be used to rewrite the action~\eqref{eq:NarainActionLR} as
\equ{ \label{eq:NarainCovariantActionLRSplitted} 
S= \int \dfrac{\d^2\gs}{2\pi}
\Big[ 
- \dfrac 1{\sqrt 2}\, \der_1 Y^T \cE^T
\Big(  \cP_\text{L}\, \cE\, \bder Y + \cP_\text{R}\, \cE\,\der Y \Big)
+ \dfrac 12\,(\der_1 x)^2 - \dfrac 12\,(\der_0 x)^2 
- \gPs^T \der \gPs
- \gps^T \der \gps
\Big]~. 
}
Following the second method of identifying the supersymmetry transformations discussed in Subsection~\ref{sc:ODDSusyWS} one obtains
\begin{subequations} \label{eq:NarainRightSusy}
\equa{ 
\gd Y  & = \ge\, \cE^{-1} \cP_\text{R} \cV\, \gPs = \ge 
\pmtrx{ e^{-1} \\[1ex] -e^T - C e^{-1} \\[1ex] -\ga_{16}^{-1}Ae^{-1} } \gPs~, 
\quad\text{where}\quad 
\cP_\text{R} \cV = \cV = \sqrt{2}\, \pmtrx{ \Id_D \\[1ex] 0 \\[1ex] 0}
~, 
\\[1ex] 
\gd \gPs &= -\dfrac{1}{\sqrt 2}\, \ge\, \cV^T \cP_\text{R} \cE\, \der_1 Y
= - \dfrac 1{\sqrt 2} \ge 
\pmtrx{ e+e^{-T} C^T & - e^{-T}  & e^{-T}A\ga_{16}} \der_1 Y~,
}
\end{subequations} 
since the inverse generalized vielbein is given by~\eqref{eq:InvNarainVielbein}. Again, the closure of the supersymmetry transformations on the fields $\gps$ and $Y$ can be expressed as 
\equ{
[\gd_1, \gd_2] \gPs = 2\sqrt{2}\, \ge_1\ge_2\, \der_1 \gPs
\qand
[\gd_1, \gd_2] Y = 2\sqrt{2}\, \ge_1\ge_2\, \dfrac{\Id - \hcZ}2\, \der_1 Y~. 
}

\subsection{One-Loop Partition Function}

The properties~\eqref{eq:Z2GradingProperties} imply, that the determinant relation~\eqref{eq:DetGeneralizedODDMetrics} changes to 
\equ{ \label{eq:DetGeneralizedODD16Metrics}
\det{}_{2D} \big[ a\, \hget + b\, \hcH \big] = (-)^D 
 \big(a+b\big)^{D+16}\, \big(a - b\big)^D~, 
}
which is again independent of all moduli; in this case $G, B, A$\,. This is derived in appendix~\ref{sc:GenMetricIndependence}, see~\eqref{eq:DetRelationNarain} with $\cA=a\, \Id$, $\cB=b\, \Id$. 
The full one--loop partition function is then given by 
\equ{
\label{eq:HetPartition}
Z_\text{full}(\gt, \bgt) =  Z_\text{Mink}(\gt, \bgt)\, Z_\text{Ferm}(\bgt)\, Z_\text{Narain}(\gt, \bgt)~, 
}
where  
\begin{subequations} \label{eq:HetPartitions}
\equa{  \label{eq:MinkowskiPartition}
Z_\text{Mink}(\gt, \bgt) & = \dfrac 1{\gt_2^{d/2-1}}\, \left|\dfrac 1{\get^{d-2}(\gt)}\right|^2~, 
\\[2ex] \label{eq:FermionPartition} 
Z_\text{Ferm}(\bgt) & = \frac 12\, \dfrac 1{\bget^4(\bgt)}\, \sum_{s,s'=0}^{1} e^{\gp i\, s's}\, \bgth_4\brkt{\frac {1-s}2 e_4}{\frac{1-s'}2 e_4}(\bgt)~, 
\\[2ex] \label{eq:NarainPartition} 
Z_\text{Narain}(\gt, \bgt) & = \dfrac{1}{\bget(\bgt)^{D}\, \get(\gt)^{D+16}}\,  \sum_{N\in \Intr^{2D+16}}
e^{ 2\pi i\, \frac 12\, N^T \big( \gt_1\, \hget + i\gt_2\, \cH\big) N}~.
}
\end{subequations} 
In the Narain partition function~\eqref{eq:NarainPartition} there is only the sum over $N$ but not over $N'$, just like in the doubled partition function~\eqref{eq:PartitionODD}: A similar phase factor, as discussed there, has been included to ensure that the sum over $N'$ just results in an irrelevant constant infinite factor.

The partition function of the non--compact bosons representing Minkowski space in light cone gauge is modular invariant by itself: 
\equ{
Z_\text{Mink}(\gt+1, \bgt+1) = Z_\text{Mink}(\gt, \bgt)
\qand
Z_\text{Mink}\Big(\dfrac{-1}{\gt}, \dfrac{-1}{\bgt}\Big) = Z_\text{Mink}(\gt, \bgt)~. 
}
Given that $Z_\text{Ferm}(\bgt) = \overline{Z_4\brkt{0}{0}(\gt)}$ defined in~\eqref{eq:LMpartitionFun} (with $d = 2\gn = 4$), it follows from~\eqref{eq:ModCovTransPartition} that 
\equ{
Z_\text{Ferm}( \bgt+1) = e^{-2\pi i\, \frac 13}\, Z_\text{Ferm}(\bgt)
\qand
Z_\text{Ferm}\Big( \dfrac{-1}{\bgt}\Big) = Z_\text{Ferm}(\bgt)~. 
}
Finally, the modular properties of the Narain partition function read
\equ{\label{eq:ModTransNarain} 
Z_\text{Narain}(\gt+1, \bgt+1) = e^{-2\pi i\, \frac 23}\, Z_\text{Narain}(\gt, \bgt)
\qand
Z_\text{Narain}\Big(\dfrac{-1}{\gt}, \dfrac{-1}{\bgt}\Big) = Z_\text{Narain}(\gt, \bgt)~. 
}
Hence, the full partition function is modular invariant. 

The modular transformations of $Z_\text{Mink}$ and $Z_\text{Ferm}$ are rather standard and follow directly from the properties of the Dedekind and the theta--functions recalled in Appendix~\ref{sc:ModFunTrans}. The modular transformations of the Narain partition function as given here are less standard and therefore it is instructive to explain them in more detail: 

The first relation~\eqref{eq:ModTransNarain} follows upon using that under $\gt_1 \mapsto \gt_1+1$ the Narain lattice part in~\eqref{eq:NarainPartition} is invariant up to a factor $\exp\{2\pi i\, \frac 12 N^T \hget N\}$. Given the form~\eqref{eq:ODD16Metric} of $\hget$, it follows that $N^T\hget N \in 2\,\Intr$ so that this factor is simply equal to unity. Hence, only each of the 16 factors $\get(\gt)$ give rise to the non--trivial phase in the first relation~\eqref{eq:ModTransNarain}; see the modular transformation~\eqref{eq:ModTransDedekind} of the Dedekind function.  

The second equation in~\eqref{eq:ModTransNarain} results from a Poisson resummation, which can be cast in the form
\equ{ 
\sum_{N\in\Intr^{2D+16}} \gd(Y-N) = 
\sum_{M\in\Intr^{2D+16}} e^{2\pi i\, M^T\, \hget\, Y}~, 
} 
since both $\hget$ and $\hget^{-1}$ are integral as $|\!\det \hget |=1$. Applying this to the Narain lattice sum gives 
\equ{
\sum_{N} e^{2\pi i\, \frac 12\, N^T\big(\gt_1\, \hget + i\gt_2\, \hcH\big) N}
= \int \d Y e^{-2\pi \, \frac 12\, Y^T\big(\gt_2\, \hcH-i\gt_1\, \hget \big) Y} 
\sum_M e^{-2\pi i\, \frac 12\, M^T\, \hget \big(\gt_1\, \hget + i\gt_2\, \hcH\big)^{-1} \hget\, N}~, 
}
upon shifting the integration variables $Y$. The integral over $Y$ can be evaluated to 
\equ{ 
 \int \d Y e^{-2\pi \, \frac 12\, Y^T\big(\gt_2\, \hcH-i\gt_1\, \hget \big) Y} 
= \dfrac 1{\big[ \det\big(\gt_2\, \hcH-i\gt_1\, \hget\big)\big]^{1/2}} 
= \dfrac 1{|\gt|^D (-i\gt)^8}
}
using~\eqref{eq:DetGeneralizedODD16Metrics} and assuming that $D$ is even. The sum over $M$ is identical to the original sum over $N$ except for the kernel 
\equ{
-\hget \big(\gt_1\, \hget + i\gt_2\, \hcH\big)^{-1} \hget = 
-\dfrac {\gt_1}{|\gt|^2}\, \hget + i \dfrac{\gt_2}{|\gt|^2}\, \hcH~. 
}
This result can be obtained by the virtue that $\hget^{-1} \hcH$ squares to the identity~\eqref{eq:Z2GradingProperties}. Here the modular transformations~\eqref{eq:ModTransReIm} of the real $\gt_1$ and the imaginary parts $\gt_2$ of $\gt$ can be recognized.  
Hence, using the modular properties~\eqref{eq:ModTransDedekind} of the Dedekind function the second equation in~\eqref{eq:ModTransNarain} follows (when reading it from right to left).

\section{Narain Orbifolds}
\label{sc:NarainOrbifolds} 
\setcounter{equation}{0}

\subsection{General Construction of Narain Orbifolds}

\subsubsection*{Point Group and Orbifold Action}

Orbifolds are obtained by enforcing invariance of the theory under the action of a finite point group $\mathbf{P}$. 
The total number of elements in the point group is denoted by $|\mathbf{P}|$. The action~\eqref{eq:TseylinNarainAction} is preserved by orbifold transformations of the form: 
\equ{ \label{eq:OrbiActions}
\gth [y] = \hcR_\gth\, y + 2\pi\, V_\gth~, 
}
on {\em any} vector $y$ (not just the Narain coordinate fields $Y$), provided that 
\equ{ \label{eq:OrbifoldTwistConditions} 
\hcR_\gth^T \, \hcH \, \hcR_\gth = \hcH~, 
\quad 
\hcR_\gth^T \, \hget \, \hcR_\gth = \hget
\qand
\hcR_\gth\, N \in \Intr^{2D+16}~, 
}
for all $\gth\in\mathbf{P}$ and $N\in\Intr^{2D+16}$. The final condition results from the requirement, that the doubled torus periodicities~\eqref{eq:NarainLattice} need to be respected. The last two conditions imply that 
\equ{ 
\hcR_\gth \in \text{O}_{\hget}(D,D+16;\Intr)~. 
}
The identity element $\gth=1$ has $\hcR_1 = \Id$ and $V_1=0$. The composition of point group transformations is defined as
$\gth' \gth[Y] = \gth'\big[ \gth[Y]\big]$\,, hence 
\equ{ \label{eq:OrbifoldCompositionRules} 
\hcR_{\gth' \gth} =  \hcR_{\gth'}\, \hcR_{\gth}
\qand 
V_{\gth'\gth} = \hcR_{\gth'}\, V_{\gth} + V_{\gth'}~. 
}
If the order of the point group element $\gth \in \mathbf{P}$ is denoted as $|\gth|$, it follows that $\gth^{|\gth|} [Y] \sim Y$. Writing this out explicitly, leads to 
\equ{ \label{eq:OrbiActionConditions}
\hcR_\gth{}^{|\gth|} = \Id
\qand 
V_{\gth^{|\gth|}} = |\gth|\, \cP{}_\gth^\parallel\, V_\gth \in \Intr^{2D+16}~, 
}
where the projectors $\hcP{}_\gth^\parallel$, on the directions in which $\hcR_\gth$ act trivially $\hcR_\gth\, \hcP{}_\gth^\parallel = \cP{}_\gth^\parallel\, \hcR_\gth = \hcP{}_\gth^\parallel$\,, are defined as 
\equ{ \label{eq:InvariantSubspaceProjector} 
\hcP{}_\gth^\parallel = \dfrac1{|\gth|} \sum_{j=0}^{|\gth|-1} \hcR_\gth{}^j
\qand 
\hcP{}_\gth^\perp = \Id - \hcP{}_\gth^\parallel~.
} 
Since the action is invariant under a shift $Y_0$ of the origin of the coordinate system defined by $Y$, the vectors $V_\gth$ may be redefined as
\equ{ 
V_\gth \mapsto V_\gth' = V_\gth - (\Id - \hcR_\gth) Y_0~. 
}
These transformation may be used to set a certain number of components of the vectors $V_\gth$ to zero; but these are never in the directions in which $\hcR_\gth$ act trivially.

\subsubsection*{Space Group}

When the point group $\mathbf{P}$ is combined with the lattice identifications the so--called space group $\mathbf{S}$ is obtained. The space group is parameterized by element $g=(\gth, N)\in \mathbf{S}$ where $\gth \in \mathbf{P}$ and $N\in\Intr^{2D+16}$. The space group is generated by the elements 
\equ{
(\gth,0)[y] = \gth[y] = \hcR_\gth\, y + 2\pi\, V_\gth
\qand
(1,N)[y] = y + 2\pi\, N~. 
}
Hence, a general element of the space group acts as: 
\equ{
g[y] = (\gth, N)[y] = (1,N)(\gth,0)[y] = 
\hcR_\gth\, y + 2\pi\, (V_\gth + N)~. 
}
Notice that the order in which these actions are applied is important here, since
\equ{
(\gth,0)(1,N)[y] = 
\hcR_\gth\, y + 2\pi\, (\hcR_\gth\, N + V_\gth)
}
does not equal $(\gth,N)[Y]$ defined above. Indeed, since the space group acts on any vector $y$ not just the Narain coordinate fields $Y$, the general composition rule reads
\equ{ 
g'g = (\gth'\gth; \hcR_{\gth'} N+N')~, 
}
where the composition of the orbifold actions~\eqref{eq:OrbifoldCompositionRules} has been used.

The space group $\mathbf{S}$ is in general non--commutative (even if the point group $\mathbf{P}$ is). Two space group elements $g=(\gth,N)$ and $g'=(\gth',N')$ only commute if
\equ{ \label{eq:CommutingSpaceGroupElements} 
\hcR_{\gth'}\, \hcR_{\gth} = \hcR_{\gth}\, \hcR_{\gth'}
\qand 
(\Id - \hcR_{\gth'}) (N + V_{\gth})  = (\Id - \hcR_{\gth}) (N' + V_{\gth'})~. 
}

\subsubsection*{Narain Orbifold Fixed Points}

Fixed points $Y_\text{fix}$ of a space group element $g\in\mathbf{S}$ are defined by the condition 
\equ{
Y_\text{fix} \stackrel{!}{=} g[Y_\text{fix}] = \hcR_\gth\, Y_\text{fix} + 2\pi(V_\gth + N)~.
}
By bringing the first term on the right to the left--hand--side shows that this equation only has solutions provided that $\hcP{}^\parallel_\gth(V_\gth+N)=0$, hence the fixed point condition becomes
\equ{ \label{eq:FixedPointCondition} 
\big(\Id - \hcR_\gth \big) Y_\text{fix}  =  2\pi\, \hcP{}^\perp_\gth (V_\gth +N)~, 
}
which can be solved by the expression: 
\equ{ 
Y_\text{fix} = 2\pi\, 
\sum_{j=0}^{|\gth|-1} \dfrac{j}{|\gth|}\, \Big( \hcP{}^\parallel_\gth - \hcR_\gth{}^j\Big) (V_\gth +N)~. 
} 
To derive that this determines the fixed points, first observe that while the matrix $\Id - \hcR_\gth$ is not invertible, it is invertible on the subspace defined by $\hcP{}^\perp_{\gth}$: Only one of the eigenvalues, $\exp(2\pi i\, j/|\gth|)$ for $j = 0, \ldots, |\gth|-1$, of $\hcR_\gth$ is equal to unity, but that one is excluded by this projection operator. On the corresponding subspace one can show that 
\equ{ \label{eq:FixedPointGenerator} 
\Big( \Id - \hcR_\gth\Big)^{-1} \hcP{}^\perp_\gth = 
\sum_{j=0}^{|\gth|-1} \dfrac{j}{|\gth|}\, \Big( \hcP{}^\parallel_\gth - \hcR_\gth{}^j\Big)~, 
}
by making a general power expansion in $\hcR_\gth$ of the left--hand--side and multiplying this by $\Id-\hcR_\gth$ and requiring that this equals $\hcP{}^\perp_\gth$. This determines the expansion uniquely up to adding an arbitrary constant to $\hcP{}^\parallel_\gth$. This constant is fixed by requiring that projecting with $\hcP{}^\parallel_\gth$ should give zero.

\subsubsection*{Orbifold Compatible Residual Gauge Symmetry}

Not all fixed points identified by this equation are physical. Only those fixed points that cannot be removed by the residual gauge transformations~\eqref{eq:DoubledCoordConstantShifts} uncovered in Section~\ref{sc:DualityCovariantWorldsheet} are physically distinct. 
Given the boundary conditions for the doubled coordinate fields, the constant gauge parameters $\gL_0$ are subject to 
\equ{ \label{eq:GaugeTransFixedPoints}
\Big( \Id - \hcR_\gth \Big) \gL_0 = 2\pi\, L~, 
}
where $L \in \Intr^{2D+16}$. In order that this gauge symmetry is compatible with the orbifold action, the constant zero model gauge parameters have to be modified to
\footnote{
The author is embedded to H.P.\ Nilles for pointing out that the previous discussion at this point did not apply to standard symmetric orbifolds. 
} 
\equ{ \label{eq:OrbiCompResGaugeTrans} 
\gL_0 = \gL_\text{fix} +  \hcP^\parallel_\mathbf{P} \pmtrx{ \gm_0 \\ 0 \\ 0 }~, 
\qquad 
\hcP^\parallel_{\mathbf{P}} = \frac 1{|\mathbf{P}|} \sum_{\gth\in\mathbf{P}} \hcR_\gth~,  
}
where $\gL_\text{fix}$ are particular solutions to \eqref{eq:GaugeTransFixedPoints} which are non--vanishing only if $L\neq 0$ and have to be of the form $(\gl_0, 0, 0)$. 
In addition to the particular solution the homogeneous part of \eqref{eq:GaugeTransFixedPoints} admits an arbitrary solution using $\cP^\parallel_{\mathbf{P}}$, that projects on the invariant subspace of the whole point group $\mathbf{P}$. 
For symmetric orbifolds the second term in \eqref{eq:OrbiCompResGaugeTrans} is trivial (or points only in none orbifolded directions) and $\gL_\text{fix}$ define the fixed points on the original torus which can be removed by the gauge symmetry. For asymmetric orbifolds $\gL_\text{fix}$ does not lie in the subspace $(\Real^D, 0,0)$ and is therefore not admissible.


\subsection{Construction of an Orbifold Compatible Generalized Metric}
\label{sc:ConstructionOrbifoldCompatibleMetric}

The first condition in~\eqref{eq:OrbifoldTwistConditions} is not so much a condition on the integral representation matrices $\hcR_\gth$\,, but rather an existence condition of an appropriate generalized metric $\hcH$ and therefore the Narain orbifold itself. Suppose a real representation $R_\gth \in \mathrm{O}(D;\Real)\times\mathrm{O}(D+16;\Real) \subset \mathrm{O}_\get(D,D+16;\Real)$, i.e.\
\equ{ \label{eq:PropertiesRgth} 
R_\gth^T R_\gth = \Id
\qand 
R_\gth^T \get R_\gth = \get~,
}
where $\get$ is given in~\eqref{eq:MinkowskiDD16Metric}, 
of the point group $\mathbf{P}$ has been constructed. Hence, these matrices may be displayed as 
\equ{ \label{eq:BlockDD16Matrices}
R_\gth = \pmtrx{ R_{\text{R}\,\gth} & 0 \\[1ex] 0 & R_{\text{L}\, \gth} }~, 
\quad
R_{\text{R}\,\gth} \in \mathrm{O}(D;\Real) 
\qand
R_{\text{L}\, \gth} \in \mathrm{O}(D+16;\Real)~. 
}

Next, consider the matrix $\cM_\text{inv}$ constructed as 
\equ{ \label{eq:InvM}
\cM_\text{inv} = \dfrac 1{|\mathbf{P}|} \sum_{\gth\in\mathbf{P}} 
R_\gth^{-1}\, \cM_0\, \hcR_\gth~, 
}
from a generic real $2D+16\times 2D+16$--matrix $\cM_0$, 
such that $\cM_\text{inv}$ is invariant under the full point group $\mathbf{P}$. Indeed, under any element $\gth$ of the point group, this generalized vielbein $\cM_\text{inv}$ maps to itself:
\equ{ 
\big(R_\gth\big)^{-1}\, \cM_\text{inv}\, \hcR_\gth = 
 \dfrac 1{|\mathbf{P}|} \sum_{\gth'\in\mathbf{P}} 
R_\gth^{-1} R_{\gth'}^{-1}\, \cM_0\, \hcR_{\gth'} \hcR_{\gth}
=
 \dfrac 1{|\mathbf{P}|} \sum_{\gth''\in\mathbf{P}} 
R_{\gth''}^{-1}\, \cM_0\, \hcR_{\gth''} =
\cM_\text{inv}~, 
}
since $\gth'\in\mathbf{P}$ and $\gth''=\gth'\gth \in\mathbf{P}$ both label the full point group $\mathbf{P}$\,. Thus the matrix $\cM_\text{inv}$ would be a candidate for an invariant vielbein $\cE_\text{inv}$, from which the Narain moduli $G,B,A$ could be read off, provided that one chooses it such that it satisfies~\eqref{eq:PropertyGeneralizedVielbein}. However, solving this non--linear constraint can prove difficult and is in fact not necessary to determine these moduli. Indeed, define the following invariant $\Intr_2$--grading 
\equ{ \label{eq:InvZ2Grading} 
\hcZ_\text{inv} = \cM_\text{inv}^{-1}\,\get\, \cM_\text{inv}~.
}
It squares to the identity, because $\get$ does, and it is point group invariant: 
\equ{
\hcR_\gth^{-1}\, \hcZ_\text{inv}\, \hcR_\gth = 
\hcR_\gth^{-1}\, \cM_\text{inv}^{-1}\,R_\gth\, R_\gth^{-1}\,\get\, R_\gth\,  R_\gth^{-1} \cM_\text{inv}\, \hcR_\gth = \cM_\text{inv}^{-1}\,\get\, \cM_\text{inv} = \hcZ_\text{inv}~,  
}
because $R_\gth^{-1} \get R_\gth = \get$\,, 
since $R_\gth$ satisfies~\eqref{eq:PropertiesRgth}. 
An invariant generalized metric can then be obtained from this by the simple relation
\equ{ \label{eq:InvGeneralizedMetric}
\hcH_\text{inv} = \hget \,\hcZ_\text{inv}~. 
}
This form ensures that $\hcH_\text{inv}$ satisfies the quadratic constraint~\eqref{eq:Z2GradingProperties} automatically, since $\hcZ_\text{inv}$ squares to the identity. The other constraint that $\hcH_\text{inv}$ is symmetric is not implemented and hence needs to be enforced afterwards. From the generalized metric~\eqref{eq:InvGeneralizedMetric} or the $\Intr_2$--grading~\eqref{eq:InvZ2Grading} the Narain moduli $G,B,A$ can be read off using~\eqref{eq:GeneralizedNarainMetric} or~\eqref{eq:Z2GradingExplicitly}: If one starts with a completely generic $\cM_0$\,, this procedure determines both the values of the frozen moduli as well as the unconstraint ones in the form of free parameters.

In order to obtain an appropriate representation $R_\gth \in \mathrm{O}(D;\Real)\times\mathrm{O}(D+16;\Real)$  one may proceed as follows: 
\enums{ 
\item 
Block diagonalize all elements $\hcR_\gth$ of the point group $\mathbf{P}$ simultaneously over the real numbers using a similarity transformation $U\in \text{GL}(2D+16;\Real)$. 
\item 
By additional permutations bring all elements $U^{-1} \hcR_\gth U$ simultaneously in a right ($D\times D$) and a left ($D+16\times D+16$) form given in~\eqref{eq:BlockDD16Matrices}. 
}
This procedure always works, since the point group $\mathbf{P}$ has finite order and hence the matrices $\hcR_\gth$ lie in the compact part of $\mathrm{O}_{\hget}(D,D+16;\Real)$. 
However, this does not necessarily lead to a valid generalized metric $\hcH$: It might happen that $\cM_\text{inv}$ is not invertible and the $\Intr_2$--grading~\eqref{eq:InvZ2Grading} cannot be defined. Secondly, it might happen that the metric $G$ read off from~\eqref{eq:InvGeneralizedMetric} using the explicit form~\eqref{eq:GeneralizedNarainMetric} is not positive definite. This means that the distribution of the blocks chosen in the second step is not appropriate and another distribution should be considered. 

The procedure outlined here may seem to be somewhat awkward in light of the issues that the generalized metric $\hcH$ is not automatically symmetric or that the metric $G$ is not necessarily positive definite. However, these are minor concerns as compare to the situation before: In~\cite{GrootNibbelink:2017usl} the moduli were constraint by a coupled set of Riccati matrix equations (see (5.17) in that reference) for which the very existence of solutions is unclear and the explicite determination of such solutions is a highly non--trivial task. Therefore, that here a concrete procedure is unfolded might be considered as a definite step forward in the construction of asymmetric orbifolds.  In Appendix~\ref{sc:T-folds} examples are given in which this procedure has been executed to obtain the most general generalized metrics for certain asymmetric orbifolds.

In the following it is assumed that the generalized metric $\hcH$, the vielbein $\cE$ and the $\Intr_2$--grading $\hcZ$ are invariant under the orbifold action and hence the subscript $inv$ is dropped. In particular, the generalized vielbein $\cE$ satifies 
\equ{ \label{eq:OrbifoldTwistConditionsVielbein} 
\big(R_\gth\big)^{-1} \cE\, \hcR_\gth = \cE~,
}
for all $\gth\in\mathbf{P}$\,.

\subsection{Dimension of the moduli space of a Narain Orbifold}

As was argued in~\cite{GrootNibbelink:2017usl}, the dimension of the untwisted moduli space of a Narain orbifold is given by 
\equ{ \label{eq:DimModuliSpaceNarainOrbifold} 
\text{dim}\big[ \fM_\mathbf{P} \big] = \dfrac 1{|\mathbf{P}|}
\sum_{\gth\in\mathbf{P}} \gch_R(\gth)\, \gch_L^*(\gth)~, 
\quad
\gch_\text{L/R}(\gth) = \tr\big[ R_{\text{L/R}\, \gth}\big] =  
 \tr\Big[\dfrac{\Id\pm \get}{2} R_\gth \Big]~. 
}
In particular, when either the left-- or right--moving twist representation is trivial while the other is not, the dimension of the moduli space is zero. Hence, this result may be used to confirm if all unconstraint moduli have been identified.

\subsection{Narain Orbifold Worldsheet Torus Boundary Conditions} 

On the worldsheet the Narain coordinates $Y$ are periodic up to actions of the space group $\mathbf{S}$: 
\begin{subequations} \label{eq:NarainOrbifoldBoundaryConditions}
\equa{
Y(z+1) &= \hcR_{\gth}\, Y(z) + 2\pi\, (N + V_\gth)
\qand  \\[1ex]
Y(z-\gt) &= \hcR_{\gth'}\, Y(z) + 2\pi\, (N' + V_{\gth'})~. 
}
\end{subequations}
The boundary conditions of the superpartners $\gps$ of the non--compact coordinate fields $x$ are the standard ones: 
\begin{subequations} \label{eq:OrbiBoundaryConditionsRMFermions} 
\equ{
\gps(z+1) = (-)^s\,\gps(z)
\qand
\gps(z-\gt) = (-)^{s'}\, \gps(z)~,  
}
where the spin structures $s, s'=0,1$ have been included.
The boundary conditions for the right--moving worldsheet fermions $\gPs$ have to be compatible with the worldsheet supersymmetry transformations~\eqref{eq:NarainRightSusy} linking them with the Narain coordinates $Y$\,. Using~\eqref{eq:OrbifoldTwistConditionsVielbein} one infers that their boundary condtions read
\equ{
\gPs(z+1) = (-)^s\, R_{\text{R}\,\gth}\, \gPs(z)
\qand
\gPs(z-\gt) = (-)^{s'}\, R_{\text{R}\,\gth'}\, \gPs(z)~. 
}
\end{subequations} 

Since, the matrices $R_{\text{R}\,\gth}$ and $R_{\text{R}\,\gth'}$ commute, see~\eqref{eq:CommutingSpaceGroupElements}, they can be simultaneously diagonalized over the {\em complex} numbers to 
\equ{ \label{eq:DiagonalizationRR}
S_{\text{R}\,\gth,\gth'}^{-1}\,R_{\text{R}\,\gth}\,S_{\text{R}\,\gth,\gth'}^{\phantom{-1}} = e^{2\pi i\, \mathbf{v}_{\text{R}\,\gth}}
\qand 
S_{\text{R}\,\gth,\gth'}^{-1}\,R_{\text{R}\,\gth'}\,S_{\text{R}\,\gth,\gth'}^{\phantom{-1}}  = e^{2\pi i\, \mathbf{v}_{\text{R}\,\gth'}}~, 
}
where the $D\times D$--matrix $\mathbf{v}_{\text{R}\,\gth}$ is associated to the {\em real} twist vector $v_{\text{R}\,\gth} = (v_{\text{R}\,\gth\,1}, \ldots, v_{\text{R}\,\gth\,D})$, as defined in~\eqref{eq:MatrixTwistVector}. For elements that involve rotations in two dimensions, the eigenvalues come in complex conjugate pairs.
Target space spinors are admitted only if, the orbifold twists preserve orientation hence their determinants have to be equal unity.

Only the real twist vector $v_{\text{R}\, \gth}$ is relevant to determine how many target space supersymmetries are perserved in the non--compact dimensions. No supersymmetries are preserved if in succession:  
\begin{itemize}
\item an odd number of entries of $v_{\text{R}\, \gth}$ are non--zero;
\item not all non--zero entries can be chosen in opposite signed pairs;
\item there is only one such pair. 
\end{itemize}
If none of  these three conditions are satisfied, a complex twist vector $v_{\text{R}\Cplx\,\gth}$ can be associated to the real twist vector by taking one of the two entries of the opposite signed pairs augmented by a number of zeros such that this complex vector has four entries. By the final condition it follows that at least two entries are non--zero. If there is a choice of these entries such that 
\equ{ \label{eq:SusyCondition} 
\sfrac12 e_4 \cdot v_{\text{R}\Cplx\,\gth} 
\equiv 0~, 
}
at least $\cN=1$ supersymmetry is preserved in the $d$ non--compact dimensions.

Because of the inhomogeneous terms, solving the boundary conditions~\eqref{eq:NarainOrbifoldBoundaryConditions} of the Narain coordinates $Y$ is more involved. To this end the following ansatz is made 
\equ{ \label{eq:AnsatzY} 
Y(z) = Y_\text{c}(z)+ Y_\text{q}(z)~, 
\qquad
Y_\text{c}(z) = 2\pi(y_0 + p\, z + \bp\,\bz) 
}
where $y_0, p, \bp$ are constant vectors and $Y_\text{q}(z)$ is assumed to satisfy the homogeneous boundary conditions 
\equ{ \label{eq:QuantumBoundaryConditionsY} 
Y_\text{q}(z+1) = \hcR_\gth\, Y_\text{q}(z)
\qand
Y_\text{q}(z-\gt) = \hcR_{\gth'}\, Y_\text{q}(z)~, 
}
where the twist matrices can be written as complex exponentials 
\equ{
\hcR_{\gth} = e^{2\pi i\, \mathbf{v}_{\gth}}
\qand 
\hcR_{\gth'} = e^{2\pi i\, \mathbf{v}_{\gth'}}~, 
}
in terms of the matrices $\mathbf{v}_{\gth}$ and $\mathbf{v}_{\gth'}$ which mutually commute. 

The zero modes contained in $Y_\text{c}(z)$ can be treated in the following fashion: Inserting the ansatz~\eqref{eq:AnsatzY} in the boundary conditions~\eqref{eq:NarainOrbifoldBoundaryConditions} leads to the requirements 
\begin{subequations} \label{eq:ConditionsZeroModes} 
\equ{ 
\hcR_{\gth} p = \hcR_{\gth'} p = p~, \qquad 
\hcR_{\gth} \bp = \hcR_{\gth'} \bp = \bp~, 
\\[1ex] \label{eq:ConditionsZeroModesII} 
\big( \Id - \hcR_{\gth}\big) y_0 + p + \bp = N+ V_{\gth}
\qand 
\big( \Id - \hcR_{\gth'}\big) y_0 -\gt\,p -\bgt\,\bp = N'+ V_{\gth'}~. 
}
\end{subequations}
The first two conditions imply that both $p$ and $\bp$ lie in both the invariant subspaces of $\hcR_\gth$ and $\hcR_{\gth'}$, which can be identified by the projectors  $\hcP{}^\parallel_{\gth}$ and $\hcP{}^\parallel_{\gth'}$ given in~\eqref{eq:InvariantSubspaceProjector}: 
\equ{ \label{eq:ParallelNotation} 
p = p^\parallel = \hcP{}^\parallel p~, 
\qand 
\bp = \bp^\parallel = \hcP{}^\parallel \bp~, 
}
To avoid overcomplicating the notation, the projector on the combined invariant subspaces of $\hcR_\gth$ and $\hcR_{\gth'}$ is denoted simply by $\hcP^\parallel$ without the subscripts $\gth,\gth'$, i.e. 
\equ{ \label{eq:DoubleProjection}
\hcP{}^\parallel =\hcP{}^\parallel_{\gth,\gth'} =  \hcP{}^\parallel_{\gth}\, \hcP{}^\parallel_{\gth'}
\qand
\cP^\perp = \Id - \cP^\parallel~. 
}
Hence, the combined orbifold actions leave a Narain subtorus of dimension $D^\parallel$ inert, where: 
\equ{ \label{eq:ProjectedDimensions} 
D^\parallel = D^\parallel_\text{L} + D^\parallel_\text{R} = \tr\big[ \hcP^\parallel\big]
\qand 
D^\perp= D^\parallel_\text{L} + D^\parallel_\text{R} = \tr \big[\hcP^\perp\big]
}
gives the dimension of the complementary subspace identified by $\hcP^\perp$. These may be divided further into left-- and right--moving dimensions 
\equ{ \label{eq:ProjectedLMDimensions} 
D^\parallel_\text{L/R} = \tr\big[ \hcP^\parallel \hcP_\text{L/R}\big]
\qand 
D^\perp_\text{L/R}= \tr\big[ \hcP^\parallel \hcP_\text{L/R}\big]~.
}
using the left-- and right--moving projectors~\eqref{eq:SplittingNarainOperators}. In addition, 
\equ{ \label{eq:ProjectedLMDimRelations}
D^\parallel_\text{L} + D^\perp_\text{L} = D_\text{L} = D+16
\qand 
D^\parallel_\text{R} + D^\perp_\text{R} = D_\text{R} = D~.
}
Hence, by projecting the two equations in~\eqref{eq:ConditionsZeroModesII} by $\hcP{}^\parallel$ the vectors $p$ and $\bp$ can be determined to be given by: 
\equ{
p\, z + \bp\, \bz = 
\hcP{}^\parallel
\gf\brkt{N+V_{\gth}}{N'+V_{\gth'}}(z)~,  
}
using~\eqref{eq:LinearMode}.

Using the operator~\eqref{eq:FixedPointGenerator} two expressions for the constant vector $y_0$ can be obtained: 
\begin{subequations} \label{eq:FixedPointsOnWST}
\equa{
y_0 &= \Big( \Id - \hcR_{\gth\phantom{'}}\Big)^{-1} \hcP{}^\perp_\gth\hcP{}^\parallel_{\gth'}  \Big(N\phantom{'}+V_{\gth\phantom{'}}\Big)
+\hcP{}^\parallel_{\gth\phantom{'}} y_0'
+ \Big( \Id - \hcR_{\gth\phantom{'}}\Big)^{-1} \hcP{}^\perp_\gth\hcP{}^\perp_{\gth'}  \Big(N\phantom{'}+V_{\gth\phantom{'}}\Big)~, 
\\[1ex]
y_0 &= \Big( \Id - \hcR_{\gth'}\Big)^{-1} \hcP{}^\perp_{\gth'} \hcP{}^\parallel_{\gth\phantom{'}}\Big(N'+V_{\gth'}\Big)
+\hcP{}^\parallel_{\gth'} y_0''
+ \Big( \Id - \hcR_{\gth'}\Big)^{-1} \hcP{}^\perp_{\gth'}\hcP{}^\perp_{\gth\phantom{'}} \Big(N'+V_{\gth'}\Big)~,  
}
\end{subequations} 
where $y_0'$ and $y_0''$ are two arbitrary vectors. Using that the projectors defined in~\eqref{eq:InvariantSubspaceProjector} are complete, the first terms in these expressions are in independent directions that are undetermined from the other equations. Hence, for both expressions to agree, it follows that 
\begin{subequations}
\equa{ 
\hcP{}^\perp_{\gth'} \hcP{}^\parallel_{\gth\phantom{'}} y_0 &= 
\Big( \Id - \hcR_{\gth'}\Big)^{-1} \hcP{}^\perp_{\gth'} \hcP{}^\parallel_{\gth\phantom{'}}\Big(N'+V_{\gth'}\Big)~,
\\[1ex] 
 \hcP{}^\perp_\gth \hcP{}^\parallel_{\gth'} y_0' &=
\Big( \Id - \hcR_{\gth\phantom{'}}\Big)^{-1} \hcP{}^\perp_\gth \hcP{}^\parallel_{\gth'}  \Big(N\phantom{'}+V_{\gth\phantom{'}}\Big)
}
\end{subequations} 
and 
\(
\hcP{}^\parallel_{\gth'} 
\hcP{}^\parallel_{\gth\phantom{'}} y_0'' 
=
\hcP{}^\parallel_{\gth'} 
\hcP{}^\parallel_{\gth\phantom{'}} y_0'\,, 
\)
which is arbitrary and may be set to zero for convenience. The final two terms in the expressions~\eqref{eq:FixedPointsOnWST} are equal by virtue of the second equation in~\eqref{eq:CommutingSpaceGroupElements}: Multiplying it with both projectors $\hcP{}^\perp_{\gth}$ and $\hcP{}^\perp_{\gth'}$ and using that on the subspace defined by these projectors $\Id-\hcR_{\gth}$ and $\Id-\hcR_{\gth'}$ are invertible, this equality follows. 

Putting everything together, the ansatz~\eqref{eq:AnsatzY} for the Narain coordinates $Y$ on the worldsheet torus can be cast in the form:  
\begin{subequations} \label{eq:ClNarainModes}
\equa{ 
Y_c(z) &= 2\pi\Big( y_0 + \hcP{}^\parallel\,\gf\brkt{N+V_{\gth}}{N'+V_{\gth'}}(z)\Big)~, 
\\[1ex] 
y_0 &= 
\Big( \Id - \hcR_{\gth\phantom{'}}\Big)^{-1} \hcP{}^\perp_\gth  \Big(N\phantom{'}+V_{\gth\phantom{'}}\Big) 
+ \Big( \Id - \hcR_{\gth'}\Big)^{-1} \hcP{}^\perp_{\gth'} \hcP{}^\parallel_{\gth\phantom{'}}\Big(N'+V_{\gth'}\Big)~. 
}
\end{subequations} 

\subsection{Narain Orbifold Partition Functions}

The full orbifold partition function can be written as a sum over commuting space group elements $g, g'\in\mathbf{S}$: 
\equ{ \label{eq:FullOrbiPartitionI} 
Z_\text{full} = Z_\text{Mink}(\gt,\bgt) \, \dfrac 1{|\mathbf{P}|} \sum_{[g,g']=0} 
Z_\text{Ferm}\brkt{g}{g'}(\bgt) \, Z_\text{Lattice}\brkt{g}{g'}(\gt,\bgt)\, Z_\text{Quan}\brkt{g}{g'}(\gt,\bgt)~, 
}
Aside from the standard Minkowski space contribution, the various factors arise as follows:

Taking into account the boundary conditions~\eqref{eq:OrbiBoundaryConditionsRMFermions}, 
the fermionic partition function can be expressed as the complex conjugated of the result~\eqref{eq:LMpartitionFunReal} for real left--moving fermions in the Appendix~\ref{sc:FermionicPartitionFunctions}. The result only depends on the point group elements $\gth,\gth'\in\mathbf{P}$ not on the whole space group elements $g,g'\in\mathbf{S}$. The resulting expression is given below in~\eqref{eq:OrbiFermPartition}.

Inserting the expression~\eqref{eq:ClNarainModes} in the worldsheet action leads to a projected lattice sum 
\equ{ \label{eq:OrbiNarainLattice} 
Z_\text{Lattice}\brkt{g}{g'}(\gt, \bgt) = 
e^{2\pi i\, (V_{\gth'})^\parallel \,\hget\, N^\parallel}
\, 
e^{ 2\pi i\, \frac 12\, (N+V_\gth)^{\parallel\,T} \big( \gt_1\, \hget + i\gt_2\, \hcH\big) (N+V_\gth)^\parallel}~, 
}
using the notation introduced in~\eqref{eq:ParallelNotation}, in particular $(V_\gth)^\parallel = \hcP^\parallel V_\gth = \hcP^\parallel_{\gth,\gth'} V_\gth$ is projected on the combined invariant subspace defined by $\hcR_\gth$ and $\hcR_{\gth'}$ (and not just $\hcR_\gth$). 
As before in the Narain lattice case, the opposite phase to the one out front has been included to ensure that the partition function is independent of $N'$. An unwanted consequence of this procedure is that also no orbifold projection by summing over $\gth'$ is implemented anymore. The phase in front of~\eqref{eq:OrbiNarainLattice} corrects for this and will enforce the orbifold projection when summing over $\gth'$. On the subspace defined by $\hcP^\parallel$ two space group elements $g$ and $g'$ commute if their point group projections $\gth, \gth'\in\mathbf{P}$ commute. Hence, in light of these observations the sum over commuting space group elements in~\eqref{eq:FullOrbiPartitionI} can be replaced by a single sum over $N$ and a sum over commuting point group elements $[\gth,\gth']=0$.

Finally, the boundary conditions~\eqref{eq:QuantumBoundaryConditionsY} can be solved by the mode expansion 
\equ{ 
Y_\text{q}(z) = \gF\brkt{r\phantom{'}\Id + \mathbf{v}_{\gth\phantom{'}}}{r'\Id + \mathbf{v}_{\gth'}}(z,\bz)\, Y\brkt{r}{r'}~. 
}
Inserting this in the Narain action to evaluate the corresponding path integral gives
\equ{ 
Z_\text{Quan}\brkt{g}{g'}(\gt,\bgt) \sim 
\prod_{r,r'} \dfrac{1}{
\det\Big[ 
\hget \, \big( \gt_1\,(r \Id+\mathbf{v}_\gth) +r'\Id + \mathbf{v}_{\gth'} \big)+
\hcH\, i\gt_2\,(r \Id+\mathbf{v}_\gth)
\Big]}~.
}
Using~\eqref{eq:DetRelationNarainII} with $\mathbf{a} = r\Id+ \mathbf{v}_{\gth}$ and $\mathbf{a}' = r'\Id+ \mathbf{v}_{\gth'}$ this can be written as
\equ{ \label{eq:QuantumOrbiPartition} 
Z_\text{Quan}\brkt{g}{g'}(\gt,\bgt) \sim 
\prod_{r,r'} \dfrac{1}{
\det_\text{L}\big[  \gt\,(r \Id+\mathbf{v}_\gth) +r'\Id + \mathbf{v}_{\gth'} \big] \, 
\det_\text{R}\big[  \bgt\,(r \Id+\mathbf{v}_\gth) +r'\Id + \mathbf{v}_{\gth'} \big]
}~.
}
Since, like $R_{\text{R}\,\gth}, R_{\text{R}\,\gth'}$, the matrices $\hcR_{\gth}, \hcR_{\text{R}\,\gth'}$ can be diagonalized over the complex numbers,  the $2D+16\times 2D+16$--matrix $\mathbf{v}_\gth$ can be splitted in a right-- and left--moving part as
\equ{
\cS_{\gth,\gth'}^{-1}\,\mathbf{v}_\gth\,\cS_{\gth,\gth'}^{\phantom{-1}} 
 = \pmtrx{ \mathbf{v}_{\text{R}\,\gth} & 0 \\ 0 & \mathbf{v}_{\text{L}\,\gth} }~, 
}
and hence for the associated real twist vector $v_\gth = (v_{\text{R}\,\gth}, v_{\text{L}\,\gth})$. By worldsheet supersymmetry $v_{\text{R}\,\gth}$ is the same real twist vector as introduced below~\eqref{eq:DiagonalizationRR}. The  vector $v_{\text{L}\,\gth} =(v_{\text{L}\,\gth\,1},\ldots, v_{\text{L}\,\gth\,D+16} )$ are in general independent of those of $v_{\text{R}\,\gth}$; except for symmetric orbifolds, where they are equal to those of $v_{\text{R}\,\gth}$ augmented with 16 zeros. On the subspace defined by $\cP^\parallel$ defined in~\eqref{eq:DoubleProjection} both $v_\gth, v_{\gth'}$ are integral (which by redefintions of $r,r'$ maybe set to zero), hence give infinite product factors that can be written in terms of the Dedekind function via~\eqref{eq:ProductDedekind}. On the complementary subspace defined by $\cP^\perp$ either $v_\gth$ or $v_{\gth'}$ are non--integral, hence are of the form of chiral bosons~\eqref{eq:ChBosonPartition}.

Taking all this into account, the full partition function can be reshuffled to 
\equ{
Z_\text{full} = Z_\text{Mink}(\gt,\bgt) \, \dfrac 1{|\mathbf{P}|} \sum_{[\gth,\gth']=0} 
Z_\text{Ferm}\brkt{\gth}{\gth'}(\bgt) \, Z^\parallel\brkt{\gth}{\gth'}(\gt,\bgt)\, Z^\perp\brkt{\gth}{\gth'}(\gt,\bgt)~, 
}
with the right--moving fermionic partition function given by 
\begin{subequations} 
\equ{ \label{eq:OrbiFermPartition} 
Z_{\text{Ferm}}\brkt{\gth}{\gth'}(\bgt) = 
e^{\pi i\, \sfrac 12v_{\text{R}\,\gth}^T(v_{\text{R}\,\gth'}-e_8)}\, \frac 1{2} \sum_{s,s'=0}^{1}\, e^{\pi i\, s's}\, 
e^{2\pi i\, \sfrac{s'}4 e_8^T v_{\text{R}\,\gth}} \, 
 \frac{\left[\,\overline{\gth_8\brkt{\sfrac12 e_8 - \sfrac {s}{2} e_8- v_{\text{R}\,\gth}}{\sfrac 12 e_8 -\sfrac {s'}{2} e_8-v_{\text{R}\,\gth'}}(\gt)}\,\right]^{1/2}}{\get^{4}(\bgt)}~. 
}
Combining the projected Narain lattice sum~\eqref{eq:OrbiNarainLattice} with the contributions of~\eqref{eq:QuantumOrbiPartition} associated to the subspace defined by $\hcP^\parallel$, leads to 
\equ{ \label{eq:OrbiNarainParallelPartition} 
Z^\parallel\brkt{\gth}{\gth'}(\gt,\bgt) =
\dfrac{e^{2\pi i\, \sfrac 12\, (V_{\gth'})^{\parallel\,T}\,\hget\,(V_\gth)^\parallel}}{
\get(\gt)^{D^\parallel_{\text{L}}}\,    
\bget(\bgt)^{D^\parallel_{\text{R}}}  
} 
\sum_{N} 
e^{2\pi i\, (V_{\gth'})^\parallel \,\hget\, N^\parallel}
\, 
e^{2\pi i\, \sfrac 12\, (N+V_\gth))^{\parallel\,T}(\gt_1\, \hget + i\gt_2\, \hcH)(N+V_\gth)^\parallel}~. 
}
The remaining contributions of~\eqref{eq:QuantumOrbiPartition} give rise to
\equ{ \label{eq:OrbiNarainPerpPartition} 
Z^\perp\brkt{\gth}{\gth'}(\gt,\bgt) =
e^{\pi i\, \sfrac 12v_{\gth}^T\get(v_{\gth'}-e_{D^\perp})}\, 
 \frac{\get(\gt)^{\sfrac 12 D^\perp_{\text{L}}}}{\left[\gth_{D^\perp_{\text{L}}}\brkt{\sfrac12e_{D^\perp_\text{L}} - v_{\gth\phantom{'}\text{L}}}{\sfrac 12 e_{D^\perp_\text{L}} -v_{\gth'\text{L}}}(\gt)\right]^{1/2}}
 \frac{\get(\bgt)^{\sfrac 12 D^\perp_{\text{R}}}}{\left[\,\overline{\gth_{D^\perp_{\gth\,\gth'\text{R}}}\brkt{\sfrac12e_{D^\perp_\text{R}} - v_{\gth\phantom{'}\text{R}}}{\sfrac 12e_{D^\perp_\text{R}} -v_{\gth'\text{R}}}(\gt)}\,\right]^{1/2}}~.
}
\end{subequations} 
Some additional phase factors have been included in these expressions so as to ensure that these building block all transform covariantly under modular transformations. (Extended discussion on the vacuum phases to obtain modular invariant partition functions can be found in e.g.\ 
\cite{Antoniadis:1986rn,Kawai:1986vd,Kawai:1986ah,Senda:1986wp,Senda:1987pf,Scrucca:2001ni}.)

Applying the projector $\hcP^\parallel$ to~\eqref{eq:OrbifoldCompositionRules} shows that 
\equ{
(V_{\gth'\gth})^\parallel = (V_{\gth'})^\parallel + (V_\gth)^\parallel
\qand
(V_{\gth^{-1}})^\parallel = - (V_\gth)^\parallel~, 
}
consequently, the modular transformation rules become: 
\begin{subequations}
\equa{ 
Z_{\text{Ferm}}\brkt{\gth}{\gth'}(\bgt+1) = 
e^{-2\pi i\, \sfrac 8{24}} \, 
Z_{\text{Ferm}}\brkt{\gth}{\gth'\gth}(\bgt) 
&~,\quad
Z_{\text{Ferm}}\brkt{\gth}{\gth'}\big(\sfrac{-1}{\bgt}\big) = 
Z_{\text{Ferm}}\brkt{\gth'}{\gth^{-1}}(\bgt)~, 
\\[1ex] 
Z^\parallel\brkt{\gth}{\gth'}(\gt+1,\bgt+1) = 
e^{2\pi i\, \sfrac {D^\parallel_\text{R}-D^\parallel_\text{L}}{24}} \, 
Z^\parallel\brkt{\gth}{\gth'\gth}(\gt,\bgt) 
&~,\quad
Z^\parallel\brkt{\gth}{\gth'}\big(\sfrac{-1}{\gt},\sfrac{-1}{\bgt}\big) = 
Z^\parallel\brkt{\gth'}{\gth^{-1}}(\gt,\bgt)~, 
\\[1ex] \label{eq:ModTransZperp}
Z^\perp\brkt{\gth}{\gth'}(\gt+1,\bgt+1) = 
e^{2\pi i\, \sfrac {D^\perp_\text{R}-D^\perp_\text{L}}{24}} \, 
Z^\perp\brkt{\gth}{\gth'\gth}(\gt,\bgt) 
&~,\quad
Z^\perp\brkt{\gth}{\gth'}\big(\sfrac{-1}{\gt},\sfrac{-1}{\bgt}\big) = 
e^{2\pi i\, \sfrac {D^\perp_\text{L}-D^\perp_\text{R}}{8}} \, 
Z^\perp\brkt{\gth'}{\gth^{-1}}(\gt,\bgt)~, 
}
\end{subequations} 
provided that 
\equ{
\dfrac 12\, N^{\parallel\,T} \,\hget\, N^\parallel 
= \dfrac 12\, N^T\, \hget^\parallel\, N
\in \Intr~,
\quad
\hget^\parallel =  \hget\, \hcP^\parallel
}
for all $N\in\Intr^{2D+16}$. This implies that $\hget^\parallel$ should define a invariant metric for a Narain, e.g.\ even self--dual, lattice of dimension $D^\parallel$ for all commuting $\gth,\gth'\in\mathbf{P}$\,. For all such lattices $D^\parallel_\text{L}-D^\parallel_\text{R}$ is dividable by 8. Because of~\eqref{eq:ProjectedLMDimRelations} also $D^\perp_\text{L}-D^\perp_\text{R}$ is dividable by 8, hence the phase in the second equation of~\eqref{eq:ModTransZperp} is trivial.\footnote{Because of this, the orbifold twists are generically assumed to act only in $2D$ subspace of the Narain lattice.} In addition, \eqref{eq:ProjectedLMDimRelations} implies that the phases in the $\gt\ra \gt+1$--transformation combined become trivial.

Finally, to ensure that the full partition function encodes the correct orbifold action, the order of the orbifold elements needs to be checked on the level of the partition function. Since $\gth' = \gth' \gth^{|\gth|}$, this leads to the requirements
\equ{ \label{eq:ProperProjectionsPartition}
\dfrac{|\gth|}{2} \Big(
(V_{\gth})^{\parallel\, T} \hget\, (V_{\gth'})^\parallel 
- \dfrac 12\, (v_{\text{L}\,\gth})^{\perp\,T} (v_{\text{L}\,\gth'})^\perp
\Big) \equiv 0
\qand 
\dfrac{|\gth|}{2}\, \dfrac 12\, e_{D^\perp_\text{L}}^T (v_{\text{L}\,\gth})^\perp \equiv 0~, 
}
for all commuting $\gth,\gth'\in\mathbf{P}$\,. Here it was used that the non--vanishing entries of $v_{R\,\gth}$ and $v_{R\,\gth}^\perp$ are equal.

\subsection{Supersymmetric Symmetric Orbifolds}

The dominant part of the (heterotic) string literature concerns itself with a very special class of orbifold: Orbifolds with actions that are purely geometrical that treat the left-- and right--moving coordinate fields identically. Moreover, the actions are such that a certain amount of supersymmetry is preserved. 

In the language employed here this means that $v_{\text{L}\,\gth}$ has no non--vanishing components on the gauge directions and that the non--vanishing entries of $v_{\text{L}\,\gth}$ and $v_{\text{R}\,\gth}$ are equal and is simply denoted by $v_\gth$. Moreover, supersymmetry implies the existence of a complex structure so that the internal coordinate fields and fermions can be combined to complex entities. In that literature $v_\gth$ has thus four instead of eight components, so that certain factors of $1/2$ in front of the inner products with these quantities here should be removed when translated to the literature on supersymmetric symmetric orbifolds. (Alternatively in the formalism developed here this means that each entry appears twice with opposite signs.) Finally, the gauge shifts $V_\gth$ lie purely in the left--moving gauge directions of the lattice. In the supersymmetric orbifold literature there is a similar relation like the second relation in~\eqref{eq:ProperProjectionsPartition} for $V_\gth$. In the formalism here this condition is obsolete as it is already incorporated by the metric $\hget$ which contains the Cartan metric of E$_8\times $E$_8$\,.  

The treatment of the Wilson lines is very different in the standard orbifold literature and in the formalism of~\cite{GrootNibbelink:2017usl} and this paper. In the standard orbifold literature the (discrete) Wilson lines arise when torus lattice translations are combined with simultaneous shifts on the gauge lattice. In the formalism here the Wilson lines $A$ are part of the data encoded in the generalized metric $\hcH$ that also include the target space torus metric $G$ and the anti--symmetric tensor field background $B$.

\section{Shift Orbifolds and Lattice Refinements}
\label{sc:ShiftOrbifolds} 
\setcounter{equation}{0}

Shift orbifolds are special types of orbifolds that have trivial twist actions. This means that the orbifold action~\eqref{eq:OrbiActions} reduces to 
\equ{ 
\gth [y] = y + 2\pi\, V_\gth~,
}
satisfying
\equ{ 
V_{\gth^{|\gth|}} = |\gth|\, V_\gth \in \Intr^{2D+16}~, 
\qand 
\dfrac{|\gth|}{2} \,
V_{\gth}^{ T} \hget\, V_{\gth'} \equiv 0
}
see~\eqref{eq:OrbiActionConditions} and~\eqref{eq:ProperProjectionsPartition}, since the projection $\hcP^\parallel_\gth=\Id$ and $\hcP^\perp_\gth=0$, because the absence of any twist action. Consequently, the fermionic partition function~\eqref{eq:OrbiFermPartition} reduces to~\eqref{eq:FermionPartition} 
and~\eqref{eq:OrbiNarainPerpPartition} is equal to unity. Shift orbifolds are therefore entirely characterized by~\eqref{eq:OrbiNarainParallelPartition}.

\subsection{Single Shift Orbifolds}

A $\Intr_K$ shift orbifold is generated by a single shift $V_\gth = V$
\equ{ \label{ShiftLatticeBasis}
V = \sfrac 1{K}\, W~, 
\qquad 
W \in \Intr^{2D+16}~, 
\quad 
\sfrac 1{2K}\, W^2 \equiv 0~, 
}
where $W^2 = W^T\hget W$, characterized by a vector $W\in \Intr^{2D+16}$ whose entries of $W$ are relatively prime w.r.t.\ $K$. This leads to a refinement of the standard $\Intr^{2D+16}$ lattice to the lattice
\equ{ \label{ShiftedMomentumNarain}
\big\{
N + k\, V 
~\big|~ N\in\Intr^{2D+16}~,~~ k=0,\ldots, K-1 
 \big\}
}
subject to the orbifold projection condition  
\equ{ \label{singleShiftOrbiProjection} 
\sfrac 1K\, 
\big( 
W^T \hat\get N + k\, \sfrac 1{2K}\, W^2 
\big) \equiv 0~,  
} 
resulting from the phases in~\eqref{eq:OrbiNarainParallelPartition} that involve $\gth'$. 

This means that two things are happening to the lattice at the same time: the lattice is refined by the inclusion of a new lattice vector and make coarser by the orbifold projection condition which is kicking out certain lattice vectors. Hence, in addition to the integral shift vector $W$ there is a second integral associated vector $W'$ necessary in order to solve the orbifold projection condition~\eqref{singleShiftOrbiProjection}.

\subsubsection*{Lattice Decomposition}

To solve the projection condition~\eqref{singleShiftOrbiProjection} it is convenient if the original lattice $\Intr^{2D+16}$ could be decomposed in the directions of the integral vectors $W$ and $W'$ and the rest: 
\equ{ \label{DecompositionAnsatz} 
N = n\, W + n'\, W' + N^\perp~, 
}
where $n, n'$ are integers and $N^\perp$ is a $2D+16$-dimensional vector spanned by $2D+14$ integer vectors that are perpendicular to $W$ and $W'$ in the sense that  
\equ{ 
W^T\hat\get N^\perp = W^{\prime T}\hat\get N^\perp = 0~. 
}
By taking inner products of this ansatz with $W$ and $W'$ a linear system for $n,n'$ is obtained, which is readily solved 
\equ{ \label{eq:MatrixEq} 
\pmtrx{n \\ n'} = 
\frac1{\gd}\, \pmtrx{ W^{\prime 2} & - W^T\hat\get W' \\ -W^T \hat\get W' & W^2 } 
\pmtrx{W^T\hat\get N\\ W'\hat\get N}~,
}
where $\delta$ denotes the determinant of the system
\equ{ \label{ConstraintWWprime} 
\gd = 
W^2 \, W^{\prime 2} - (W^T\hat\get W')^2 
\stackrel{!}{=} \pm 1~. 
}
In general the solution for $n,n'$ won't be integral; only if $\gd^{-1}=\gd$, as enforced above, this is guaranteed. Hence, from now on it is assumed that the second vector $W'$ is chosen such that~\eqref{ConstraintWWprime} holds.

By performing the matrix multiplication in~\eqref{eq:MatrixEq} expressions for $n$ and $n'$ can be found which can be written as 
\equ{ 
\mtrx{
n\phantom{'} = W\phantom{'}\bW\phantom{'} \,N~, 
\\[1ex]
n' = W' \bW' \, N~,
} 
\qquad\text{where}\qquad
\mtrx{
\bW\phantom{'} = \gd\, \big( W^{\prime 2}\, W\phantom{'} - (W^T\hget W')\,W' \big)^T\hget~, 
\\[1ex] 
\bW' = \gd\, \big( W^{\phantom{'}2}\, W' - (W^T\hget W')\,W\phantom{'} \big)^T\hget~, 
}
}
are vectors conjugate to $W$ and $W'$ satisfying 
\equ{ 
\bW\phantom{'} W = \bW' W' = 1
\qand 
\bW\phantom{'} W' = \bW' W = 0~. 
}

Using these vectors and their conjugates two projections, $\hgP$ and $\hgP'$, and two nilpotent operators, $\hgX$ and $\hgX'$, can be introduced
\equ{ \label{eq:PXoperators}
\hgP\phantom{'} = W\phantom{'}\bW\phantom{'}~, 
\quad 
\hgP' = W' \bW'~,
\qquad
\hgX\phantom{'} = W\phantom{'} \bW'~, 
\quad 
\hgX' = W' \bW\phantom{'}~, 
}
defined by their actions 
\equ{
\mtrx{
\hgP\phantom{'} \, W\phantom{'} = W\phantom{'}~, 
\quad 
\hgP\phantom{'} \, W' = 0~,
\\[1ex] 
\hgP' \, W' = W'~,   
 \quad 
\hgP' \, W\phantom{'} = 0~,  
}
\qquad 
\mtrx{
\hgX\phantom{'} \, W\phantom{'} =  0~,
\quad 
\hgX\phantom{'} \, W' = W\phantom{'}~, 
\\[1ex] 
\hgX' \, W' = 0~,  
 \quad 
\hgX' \, W\phantom{'} =  W'~, 
}
}
on $W$ and $W'$, satisfying the algebra 
\equ{ \label{eq:PXalgebra} 
\mtrx{
\hgP^{\phantom{\prime}2} =  \hgX\phantom{'} \hgX' = \hgP\phantom{'}~, 
\\[1ex] 
\hgP^{\prime 2} = \hgX'  \hgX\phantom{'} = \hgP'~, 
}
\qquad
\mtrx{
 \hgP\phantom{'} \hgX\phantom{'} = \hgX\phantom{'} \hgP' = \hgX\phantom{'}~, 
\\[1ex]
 \hgP' \hgX' = \hgX' \hgP\phantom{'} = \hgX'~, 
}
\qquad 
\mtrx{
\hgX^{\phantom{\prime}2}\,~ = \hgX^{\prime 2}\,~ = 
\hgP\phantom{'} \hgP' = \hgP' \hgP\phantom{'} = 0~,
\\[1ex] 
\hgP\phantom{'} \hgX' = \hgP' \hgX\phantom{'} = 
\hgX\phantom{'} \hgP\phantom{'} = \hgX' \hgP' = 0~. 
}
}
The traces of these operators 
\equ{
\text{tr}[\hgP] = \text{tr}[\hgP'] = 1~, 
\qquad
\text{tr}[\hgX] = \text{tr}[\hgX'] = 0~, 
}
show that $\hgP$ and $\hgP'$ project on one dimensional subspaces. 

With this the subspaces parallel and perpendicular to $W$ and $W'$ can be easily identified by the projectors 
\equ{
\hgP_\parallel = \hgP + \hgP'~,
\qand 
\hgP_\perp = \Id - \hgP_\parallel~,  
}
satisfying 
\equ{
\hgP_\parallel^2 = \hgP_\parallel~, 
\quad 
\hgP_\perp^2 = \hgP_\perp~, 
\quad 
\hgP_\perp \hgP_\parallel = \hgP_\parallel \hgP_\perp = 0~. 
}
Using the multiplications~\eqref{eq:PXalgebra} it follows further that $\hgP_\parallel$ leaves all $\hgP,\hgP',\hgX,\hgX'$ inert irrespectively of whether multiplied from the left or the right, while $\hgP_\perp$ annihilates all of them. Their traces read
\equ{
\text{tr}[\hgP_\parallel] = 2~, 
\qquad
\text{tr}[\hgP_\perp] = 2D+14~.  
}
Hence, finally the quantities $n$, $n'$ and $N^\perp$ in the lattice decomposition \eqref{DecompositionAnsatz} can be determined from the lattice vector $N$ as 
\equ{
n\phantom{'} = \hgP\phantom{'} N~, 
\quad 
n' = \hgP' N
\qand 
N^\perp = \hgP_\perp N~. 
} 
Since the projection operators, $\hgP, \hgP$ and $\hgP_\perp$, are all integral and have unit determinant on the subspaces on which they project, this decomposition is invertible over the integers and $n,n'\in\Intr$ and $N^\perp$ lies in a $2D+14$--dimensional sublattice of $\Intr^{2D+16}$\,.

\subsubsection*{Construction of the Shift Orbifold Lattice}

To avoid arriving at very complicated formulae, the vector $W'$ is required to satisfy
\equ{ \label{SimpleConstraintConditions} 
W^{\prime 2} = 0~, 
\qquad 
\gd = -1~, 
\qquad 
W^T \hat\get W' = 1~. 
}
The first condition, in fact, only restricts $W^T \hat\get W' = \pm1$, which already implies that $\gd=-1$ by~\eqref{ConstraintWWprime}. By multiplying the whole vector with $-1$ the sign of this inner product can always be assumed to be positive.

The refined lattice vector can be written as 
\equ{ 
N+k\,V = n'\, W' + m \, \sfrac 1K\, W + N^\perp~, 
}
where $m = K\, n + k$ is an arbitrary unconstraint integer since $n\in\Intr$ and $k = 0,\ldots, K-1$. In addition, the constraint~\eqref{singleShiftOrbiProjection} can be solved as follows 
\equ{
\sfrac 1K \big( n\, W^2 + k\, \sfrac 1{2K}\, W^2 + n' \big) = 
m' + n\, \sfrac 1{2K}\, W^2~,   
}
by introducing an arbitrary $m' \in \Intr$, since $\equiv$ means equal up to integers and $\sfrac 1{2K}\, W^2\in \Intr$ by~\eqref{ShiftLatticeBasis}. The final term is added to ensure that the projection condition can be written entirely in terms of $m$ and $m'$ only. Indeed, by multiplying by $K$ and solving for $n'$ gives: 
\equ{
n' = K\big( m' + n\, \sfrac 1{2K}\, W^2  \big) -  \big( n\, W^2 + k\, \sfrac 1{2K}\, W^2 + n' \big)
= K\, m' - m\, \sfrac 1{2K}\, W^2~. 
}
This shows that the refined lattice vector can be cast in the form  
\equ{ 
N+k\, V = m\, \sfrac 1K\, W + m'\, K\,W' - m \, \sfrac 1{2K}\, W^2\, W' + N^\perp~, 
}
in terms of unconstraint integers $m,m'\in\Intr$\,. By introducing the integral vector $N' = m\, W + m'\, W' + N^\perp$, this  can be written as 
\equ{ \label{eq:NewLattice} 
N+k\, V = \hcT_0\, N'~, 
\quad 
\hcT_0 = 
 \widehat\gP_\perp + \sfrac 1K\, \widehat\gP_W + K\, \widehat\gP_{W'} 
- \sfrac 1{2K}\, W^2 \, \hgX'~,
}
using the operators~\eqref{eq:PXoperators}.

\subsubsection*{Modification of the Generalized Metric}

The expression~\eqref{eq:NewLattice} seems to suggest, the shift orbifold defines another Narain theory with a modified generalized metric $\hcH'$ obtained from an $\hcT\in\text{O}_{\hget}(D,D+16;\Ratl)$ transformation. Before this transformation can be identified, first it has to be investigated whether the lattice transformation~\eqref{eq:NewLattice} modifies $\hget$ and if so how to correct for that. To this end, define $\hget'$ by
\equ{
N^{\prime\, T} \hget'\, N' = (N+k\,V)^T \hget\, (N+k\,V)~. 
}
By inserting~\eqref{eq:NewLattice} here and using that 
\equ{
\hget^{-1} \hgP^T \hget = \hgP' + W^2\, \hgX'~, 
\quad 
\hget^{-1} \hgP^{\prime T} \hget = \hgP - W^2\, \hgX'~, 
\qquad 
\hget^{-1} \hgX^{\prime T} \hget = \hgX'~, 
}
it follows that $\hget'$ can be written as 
\equ{
\hget' = 
\hcM^T \hget \, \hcM~,
\quad 
\hcM =  \Id  - \sfrac 12\, W^2\, \hgX'
\in \text{GL}(2D+16;\Intr)~. 
}
In other words, in terms of the lattice defined by $N'$ the metric $\hget$ has changed to $\hget'$. Thus, by replacing $N'$ by the new integral vector $N'' = \hcM^{-1}N'$, the metric $\hget$ remains the same. This motivates to define $\hcT = \hcT_0\hcM$, which reads explicitly: 
\equ{ \label{eq:NarainMapping} 
\hcT = 
 \widehat\gP_\perp + \sfrac 1K\, \widehat\gP_W + K\, \widehat\gP_{W'} 
+\sfrac 1{2}\big(K - \sfrac 1K \big)\, W^2 \, \hgX'~. 
}
It may be verified that $\hcT \in \text{O}_{\hget}(D,D+16;\Ratl)$. The new generalized metric $\hcH'$ is given by 
\equ{ 
\hcH' = \hcT^T \hcH\, \hcT~.
}
This result shows explicitly that a shift orbifolds of a Narain theory is again a Narain theory but with redefined moduli.

\subsubsection*{Dependence on the Associated Vector}

The construction developed here determines the new moduli of the Narain lattice that results from a shift orbifold. However, the procedure depends on a somewhat arbitrary integral associated vector $W'$ only constraint to satisfy~\eqref{SimpleConstraintConditions}. Different choices for this vector lead to different generators of the same Narain lattice and hence different forms of the generalized vielbein and metric encoding the moduli. Seemingly this can have striking consequences, as the same shift orbifold can have different geometric and non--geometric interpretations. However, these different forms are related to each other by $T$--duality transformations $\text{O}_{\hget}(D,D+16;\Intr)$\,. Below some examples of this issue are discussed.

\subsection{A Simple Shift Orbifold}

As a first illustration it is instructive to consider the case in which the target space is a circle of radius $R$. Hence, the dual circle has radius $1/R$ in string units, i.e.\ the corresponding generalized vielbein reads 
\equ{
\hcE = \pmtrx{ R & 0 \\ 0 & \sfrac 1R }~. 
}
Performing an order--$K$ shift orbifold on this circle leads to the Narain mapping  
\equ{
\hcT = \pmtrx{ \sfrac 1K & 0 \\ 0 & K}~. 
}
Hence, the generalized vielbein becomes
\equ{
\pmtrx{ R' & 0 \\ 0 & \sfrac 1{R'} } = 
\hcE' = \hcE \, \hcT = 
\pmtrx{ R & 0 \\ 0 & \sfrac 1R }\, \pmtrx{ \sfrac 1K & 0 \\ 0 & K}
= \pmtrx{ \sfrac{R}{K} & 0 \\ 0 & \sfrac{K}{R} }~. 
}
This indicates that the target space circle is shrunk by a factor $1/K$ while the dual circle is stretched by $K$ at the same time.

\subsection{Geometrical Shift with a Wilson Line}

A Wilson line is often introduced in the orbifold literature as geometrical shift orbifold with an associated shift action on the gauge degrees of freedom. Concretely, consider an order--$K_i$ Wilson line on the $i$--th torus direction 
\equ{ 
W_i = \pmtrx{ \ge_i \\ 0 \\ a_i}
\qand 
W_i' = \pmtrx{ 0 \\ \ge_i \\ 0}~, 
}
where $a_i$ is a 16--component integral vector such that 
\equ{
K_i\, a_i\in\Intr
\qand 
\sfrac 1{2K_i}\, a_i^2 = \sfrac1{2K_i}\, a_i^T g_{16} a_i \equiv 0~. 
}
The $i$--th $D$--dimensional Euclidean basis vector $\ge_i$ satisfies
\equ{
\ge_i^T\ge_j = (\ge_i)_j = \gd_{ij}~.  
}
Since $W_i$ and $W_i'$ both involve $\ge_i$ but in different directions in the Narain lattice, it is ensured that we have an unique decomposition of the Narain lattice by~\eqref{DecompositionAnsatz}. The vector $W_i'$ is chosen such that the conditions in~\eqref{SimpleConstraintConditions} are fulfilled, hence the operators~\eqref{eq:PXoperators} are given by 
\equ{
\hgP_i = \pmtrx{ \gp_i & 0 & 0 \\[1ex] 0 & 0 & 0 \\[1ex] a_i \ge_i^T & 0 & 0}~, 
\quad 
\hgP_i' = \pmtrx{ 0 & 0 & 0 \\[1ex] -a_i^2\, \gp_i & \gp_i & \ge_i a_i^Tg_{16} \\[1ex] 0 & 0 & 0}~,  
\quad
\hgX_i' = 
\pmtrx{0 & 0 & 0 \\[1ex] \gp_i & 0 & 0 \\[1ex] 0 & 0 & 0}~, 
}
where $\gp_i = \ge_i \ge_i^T$ and $\gp_i^\perp = \Id_D - \gp_i$\,.
This results in the following Narain mapping
\equ{ 
\hcT_i = 
\pmtrx{
\gp_i^\perp + \sfrac 1{K_i}\, \gp_i & 0 & 0 \\[2ex]
\big(1 - \sfrac {K_i}2 - \sfrac1{2K_i}\big)\, a_i^2\, \gp_i & \gp_i^\perp + K_i\, \gp_i & (K_i-1)\, \ge_ia_i^T g_{16} \\[2ex] 
\big( \sfrac 1{K_i} -1\big)\, a_i\ge_i^T & 0 & \Id_{16} 
}~, 
}
using~\eqref{eq:NarainMapping}. By using the $\text{O}_{\hat\get}(D,D+16; \Real)$ elements given in~\eqref{eq:GeometricModuliMatrices} this can be written as 
\equ{
\hcT_i = \hcM_e(\gD e_i) \, \hcM_a(\gD a_i)~, 
\quad 
\gD e_i = \gp_i^\perp + \sfrac 1{K_i}\, \gp_i~, 
\qand 
\gD a_i = - \big( 1 - \sfrac 1{K_i} \big)\, \ga_{16} a_i\, \ge_i^T~. 
}
This reduces to the case discussed above when switching off the Wilson line and taking $D=1$.

Since shift actions on the Narain space commute, various geometrical shifts with Wilson lines can be combined. This simply leads to a product of the corresponding Narain mappings. If all $D$ dimensions are shift orbifolded, this gives 
\equ{ 
\hcT= \hcM_e(\gD e) \, \hcM_a(\gD a)~, 
\quad 
\gD e= \sum_i \sfrac 1{K_i}\, \gp_i~, 
\qand 
\gD a = - \sum_i \big( 1 - \sfrac 1{K_i} \big)\, \ga_{16} a_i\, \ge_i^T~. 
}
Since $\hcT$ is made up of the same building blocks as a Narain vielbein, it can be expressed as a Narain vielbein itself as 
\(
\hcT = \hcE(\gD e, 0, \gD a)\,.
\)
Hence, if the original model is described by the moduli $e, B, A$, the new moduli can be read off from the vielbein multiplication
\equ{ 
\hcE(e', B', A') = \hcE(e, B, A) \, \hcT = \hcE(e,B,A) \, \hcE(\gD e, 0, \gD a) 
}
using~\eqref{eq:ProductNarainVielbeins} to be 
\equ{
e' = E\, \gD e~,
\quad
B' = \gD e^TB\,\gD e + \sfrac 12 e^T A^T \gD a - \sfrac 12 \gD a^T  A e
\qand 
A' = A\, \gD e + \gD a~. 
}
Hence, in particular, if the anti--symmetric tensor $B$ and the Wilson lines $A$ initially were switched off, only a Wilson line background is introduce by this shift orbifold. When only the $B$--field was zero, but the initial Wilson lines $A$ were not, then the anti--symmetric tensor becomes switched on after this shift orbifold.

\subsection{Non--Geometric Shift with a Wilson Line}

Next, consider an order--$K_i$ Wilson line on the $i$--th dual torus direction
\equ{
\tW_i = \pmtrx{0 \\ \ge_i \\ a_i} 
\qand
\tW_i' = \pmtrx{\ge_i \\ 0 \\ 0}~, 
}
where $a_i$ satisfies the same properties as in the previous subsection. Hence, the same procedure can be followed as in that subsection leading to 
\equ{
\widetilde{\cT}_i = \hcM_e(\gD \tilde{e}_i) \, \hcM_\ga(\gD \ga_i)~, 
\quad 
\gD \tilde{e}_i = \gp_i^\perp +  {K_i}\, \gp_i~, 
\qand 
\gD \ga_i = - \big( 1 - \sfrac 1{K_i} \big)\, \ga_{16} a_i\, \ge_i^T~, 
}
where $\hcM_\ga$ is defined in~\eqref{eq:NonGeometricModuliMatrices}. 
Hence, in this case non--geometric gauge moduli are switched on. 

The reason how this comes about can be easily understood by the following observation. The starting data of this and the previous subsection are related to each other via the $T$--duality operator $\hI$ given in~\eqref{eq:FullTDualityOperator}: 
\equ{
\tW_i = \hI\, W_i
\qand 
\tW_i' = \hI\, W_i'~. 
}
This is just the statement that $T$--duality takes geometric shifts to non--geometric ones. Hence, one expects that also the change of the moduli due to the (non--)geometric shift with a Wilson line are related via the same operation. This is indeed the case, since 
\equ{ 
\widetilde{\cT}_i = \hI\, \hcT_i\, \hI~. 
}
This exemplifies that the interpretation of a certain transformation to be geometric or non--geometric depends on the duality frame chosen.

In this case it is only under certain circumstances possible to determine the new moduli using products of Narain vielbeins: If in the initial case $B=A=0$, then 
\equ{ 
\hcE(e,0,0) \, \widetilde{\cT}_i = \hcE(e,0,0)\, \hI\, \hcE(\gD e_i, 0, \gD a_i) \, \hI = \hI\, \hcE(e^{-T},0,0)\, \hcE(\gD e_i, 0, \gD a_i)\, \hI 
\non \\ = \hI\, \hcE(e^{-T} \gD e_i, 0, \gD a_i)\, \hI~. 
}
That is by going to the fully T--dual frame, where the initial vielbein is $e^{-T}$, the new moduli are 
\equ{
e' = e^{-T}\gD e_i
\qand 
A' = \gD a_i~.
}
However, as soon as either an anti--symmetric tensor or a Wilson line background was switched on already, this does not work any more. To determine the moduli for more general initial Narain backgrounds, one should use the generalized metric instead. The new generalized Narain metric $\hcH'$ obtained after performing this non--geometric shift with a Wilson line is given by 
\equ{ 
\hcH' = \widetilde{\cT}_i^T\, \hcH\, \widetilde{\cT}_i~. 
}

\subsection{Non--Geometric Shift in Two Dimensions} 

Consider the order--$K$ non--geometric shift 
\equ{ 
W = \pmtrx{ \ge_i \\ \ge_j \\ 0}
\qand
W_i' = \pmtrx{0 \\ \ge_i \\ 0 }~,
\quad 
W_j' = \pmtrx{\ge_j \\ 0 \\ 0 }~, 
}
for $i \neq j$\, and no action on the gauge degrees of freedom. Here two choices for the vector $W'$ are given distinguished by the label $i$ and $j$, which is used below as well for this purpose, to exemplify possible consequences of different choices of the associated vector $W'$ to solve the orbifold constraint.

In the case $W_i'$ is employed, one finds 
\equ{
\hgP_i = \pmtrx{ \gp_i & 0 & 0\\ \ge_j \ge_i^T & 0 & 0 \\ 0 & 0 & 0 }
\qand 
\hgP_i' = \pmtrx{ 0 & 0 & 0 \\ \ge_i \ge_j^T & \gp_i & 0 \\ 0 & 0 & 0}~. 
}
(Since $W^2 = W_i^{\prime 2} = 0$ and $\hgX_i'$ is irrelevant.) Hence the Narain mapping becomes 
\equ{ 
\hcT_i = \pmtrx{ 
\gp_i^\perp + \sfrac 1K\, \gp_i & 0 & 0 
\\[1ex] 
\big( \sfrac 1K - 1\big) \ge_j \ge_i^T + \big(K-1\big) \ge_i \ge_j^T & \gp_i^\perp + K\, \gp_i  & 0 
\\[1ex] 
0 & 0 & \Id_{16}   
}~. 
}
This can be written as 
\equ{ 
\hcT_i = \hcM_e(\gD e_i) \, \hcM_b(\gD b_{ij})~,
\quad
\gD e_i = \gp_i^\perp + \sfrac 1{K}\, \gp_i
\qand 
\gD b_{ij} = \big( 1 - \sfrac 1K\big) \big(\ge_i \ge_j^T - \ge_j \ge_i^T\big)~. 
}
Hence, performing a non--geometrical shift orbifold in two directions has the effect of switching on the anti--symmetric tensor field $B$ in those directions.

If $W_j'$ is used instead, the results are 
\equ{ 
\hgP_j = \pmtrx{ 0 & \ge_i \ge_j^T   & 0 \\ 0 & \gp_j & 0 \\ 0 & 0 & 0 }
\qand 
\hgP_j' = \pmtrx{ \gp_j & \ge_j \ge_i^T  & 0 \\0 & 0 & 0 \\ 0 & 0 & 0}~. 
}
leading to 
\equ{ 
\hcT_j = \pmtrx{ 
\gp_j^\perp + K\, \gp_j & \big( \sfrac 1K - 1\big) \ge_i \ge_j^T + \big(K-1\big) \ge_j \ge_i^T & 0 
\\[1ex] 
0 & \gp_j^\perp + \sfrac 1K\, \gp_j  & 0 
\\[1ex] 
0 & 0 & \Id_{16}   
}~, 
}
which can be written as 
\equ{ 
\hcT_j = \hM_e(\gD e_j) \, M_\gb(\gD \gb_{ij})~,
\quad
\gD e_j = \gp_j^\perp + {K}\, \gp_j
\qand 
\gD \gb_{ij} = \big( 1 - \sfrac 1K\big) \big(\ge_j \ge_i^T - \ge_i \ge_j^T\big)~. 
}
Hence, by making the choice $W_j'$ for the vector to solve the orbifold constraint, instead the interpretation of performing the same simultaneous shift in a torus direction and a shift in a different dual torus direction leads to a non--geometric $\gb$--background.

The two seemingly different situations can again be mapped to each other. Indeed, it is easy to see that 
\equ{ 
W = \hX_{ij} \hI\, W
\qand 
W_j' = \hX_{ij}\hI\, W_i'~, 
}
where $\hX_{ij}$ exchanges the directions $i$ and $j$ on the torus and the dual torus simultaneously. Applying these transformations to the moduli mapping $\hcT_i$ gives 
\equ{
\hX_{ij} \hI\, \hcT_i\, \hX_{ij}\hI = \hcT_j~. 
}
Hence, the interpretation of the moduli depends on the choice of the associated vector $W'$.

\section{Higher dimensional T--fold examples} 
\label{sc:T-folds} 
\setcounter{equation}{0}

\subsection{Right--Twisted Full T--Duality Narain Orbifolds}

Full T--duality orbifolds refer to Narain orbifolds in which the orbifold action is a T--duality transformation in all $D$ compact torus directions. Below two realizations of them are discussed: one in this subsection, one in the next. Even though their starting data are very similar, their properties differ in a number of interesting ways.

Consider the Narain orbifold with the $\Intr_2$ T--duality twist given by 
\equ{ 
\hcR = 
\pmtrx{0 & \Id_D & 0 \\ \Id_D & 0 & 0 \\ 0 & 0 & \Id_{16} }~.  
}
The associated block--diagonal twist is given by 
\equ{
R = 
\pmtrx{ -\Id_D & 0 & 0 \\ 0 & \Id_D & 0 \\ 0 & 0 & \Id_{16} }~, 
\quad\text{hence}\quad 
v_{\text{R}} = \pmtrx{\sfrac 12^D}
\qand 
v_{\text{L}} = 0~,
}
so that this T--duality orbifold only affects the right--movers. Hence, this T--fold can preserve target space supersymmetry only when $D$ is dividable by 4, i.e.\ $D=4,8$. Otherwise, all supersymmetry are necessarily broken.

The form of the generalized metric can be determined using the method outlined in Subsection~\ref{sc:ConstructionOrbifoldCompatibleMetric}. 
It follows, that the most general invariant matrix $\cM_\text{inv}$ can be taken to be 
\equ{
\cM_\text{inv} = 
\dfrac 12 \pmtrx{ \ga & -\ga & 0 \\ \gd & \phantom{-}\gd & 2 \gf \\ \gr & \phantom{-}\gr & 2\gt}~,
}
using~\eqref{eq:InvM}. However, given that all Narain moduli are stabilized, since 
\equ{ 
\dim\big[\fM_\mathbf{P}\big] = 
\sfrac 12 D(D+16) + \sfrac 12 (-D)(D+16) = 0
} 
by evaluating~\eqref{eq:DimModuliSpaceNarainOrbifold} explicitly,
a simplified choice for this matrix is sufficient: Setting $\gr=\gf = 0$, i.e. 
\equ{ 
\cM_\text{inv} = 
\dfrac 12 \pmtrx{ \ga & -\ga & 0 \\ \gd & \phantom{-}\gd & 0 \\ 0 & \phantom{-}0 & 2\gt}
\qand 
\cM_\text{inv}^{-1} = 
\pmtrx{ \phantom{-}\ga^{-1} & \gd^{-1} & 0 \\ -\ga^{-1} & \gd^{-1} & 0 \\ \phantom{-}0 & 0 & \gt^{-1}}~,
} 
assuming that $\ga, \gd$ and $\gt$ are invertible matrices. These expressions imply that 
\equ{
\hcZ_\text{inv} = 
\pmtrx{ 0 & \Id_D & 0 \\ \Id_D & 0 & 0 \\ 0 & 0 & \Id_{16} }
\qand 
\hcH_\text{inv} = 
\pmtrx{ \Id_D& 0 & 0 \\ 0 & \Id_D & 0 \\ 0 & 0 & g_{16} }~, 
}
using~\eqref{eq:InvZ2Grading} and~\eqref{eq:InvGeneralizedMetric}. Notice that these results are, in fact, independent of the precise form of $\ga, \gd$ and $\gt$ (which thus could have been chosen to be $\ga=\gd=\Id_D$ and $\gt=\Id_{16}$). This is just a reflection of the fact that this orbifold fixes all moduli. They read: $G= \Id_D$ and $B=A=0$.

The definition of a Narain orbifold is not complete without specifying the Narain shift vector. Given that the parallel and perpendicular projectors~\eqref{eq:InvariantSubspaceProjector} take the form
\equ{ 
\hcP^\parallel = 
\dfrac 12 
\pmtrx{
 \Id_D & \Id_D & 0 \\ 
 \Id_D & \Id_D & 0 \\ 
 0 & 0 & \Id_{16}
}~, 
\quad 
\hcP^\perp = 
\dfrac 12 
\pmtrx{
  \phantom{-}\Id_D & -\Id_D & 0 \\ 
 -\Id_D &  \phantom{-}\Id_D & 0 \\ 
  \phantom{-}0 &  \phantom{-}0 & 0
}~, 
\quad\text{the shift vector}\quad
V^\parallel = \pmtrx{ w_\text{R} \\ w_\text{R} \\ w_{16}}~, 
}
is parameterized by integral vectors $w_\text{R}\in \Intr^D$ and $w_{16} \in \Intr^{16}$\,. The consistency condition~\eqref{eq:ProperProjectionsPartition} takes the form 
\equ{
\dfrac 14\, \Big( 
2\, w_\text{R}^2 + w_{16}^T\, g_{16}\, w_{16} 
\Big) 
= (V^\parallel)^T \hget\, V^\parallel \equiv 0 
~,
}
since $v_{\text{L}\,\gth} = 0$\,. Since $g_{16}$ is even, this is gives a mod--two condition on the $D+16$ integers, which is easily solved by requiring that the number of odd integers is even.

The fixed points of this T--fold are determined via the equation~\eqref{eq:FixedPointCondition}
\equ{ 
\pmtrx{ 
 \phantom{-}\Id_D & - \Id_D & 0 \\ 
-\Id_D &  \phantom{-}\Id_D & 0 \\ 
 \phantom{-}0 &  \phantom{-}0 & 0 
} 
Y_\text{fix} = 2 \pi\,  
\frac 12 
\pmtrx{ 
 \phantom{-}\Id_D & - \Id_D & 0 \\ 
-\Id_D &  \phantom{-}\Id_D & 0 \\ 
 \phantom{-}0 &  \phantom{-}0 & 0 
} 
\pmtrx{ n \\ \tn \\ q }~. 
}
On the subspace defined by $\cP^\perp$ this can be solved as 
\equ{  \label{eq:FixedPointsRightTwistedTfold} 
\dfrac{Y_\text{fix}}{2\gp} = \dfrac{Y_\text{fix}^\perp}{2\gp}  = 
\frac 14  
\pmtrx{ n - \tn \\ \tn -n \\ 0 } 
= 
\frac 14  
\pmtrx{ m \\  -m \\ 0 }~, 
}
where $m^1,\ldots, m^D=0,1,2,3$ labels $4^D$ distinct points on the doubled torus. However, in light of the residual gauge symmetry not all are physical. The orbifold compatible residual gauge transformations contain the following constant gauge parameters
\equ{ 
\gL_0 = 
\frac 12 
\pmtrx{ 
\Id_D &  \Id_D & 0 \\ 
\Id_D & \Id_D & 0 \\ 
0 & 0 & \Id_{16} 
} 
\pmtrx{ \gm_0 \\ 0 \\ 0 } 
= 
\frac 12 \pmtrx{ \gm_0 \\ \gm_0 \\ 0}~. 
} 
($\gL_\text{fix}$ would be of the same form as \eqref{eq:FixedPointsRightTwistedTfold}, thus for it to be of the form $(\gl_0,0,0)$, $m=0$ and hence $\gL_\text{fix}=0$.)
Exploiting the constant gauge transformations the upper $D$ components of the fixed point may be set to zero by the gauge $\sfrac 12 \gm_0 = \sfrac 14 m$. Hence a fully gauge fixed form of the fixed points can be written as 
\equ{  \label{eq:FixedPointsRightTwistedTfoldGaugeFixed} 
\dfrac{Y^\text{g.f.}_\text{fix}}{2\gp}  =
\frac 12  
\pmtrx{ 0 \\  m \\ 0 }~, 
}
with $m^1,\ldots, m^D=0,1$, leading to $2^D$ physically distinct fixed points.

\subsection{Left--Twisted Full T--Duality Narain Orbifolds}

Next, consider another $\Intr_2$ T--duality Narain orbifold with a very similar twist as the one discussed above
\equ{ 
\hcR = 
\pmtrx{\phantom{-}0 & -\Id_D & 0 \\ -\Id_D & \phantom{-}0 & 0 \\ \phantom{-}0 & \phantom{-}0 & \Id_{16} }~, 
}
so that the associated block--diagonal twist now reads
\equ{ \label{eq:LeftNarainOrbifoldvL}
R = 
\pmtrx{ \Id_D & 0 & 0 \\ 0 & -\Id_D & 0 \\ 0 & 0 & \Id_{16} }~, 
\quad\text{hence}\quad 
v_{\text{R}} = 0 
\qand 
v_{\text{L}} = \pmtrx{\sfrac 12^D, 0^{16}}~,
}
In this case the action on the right--movers is trivial. Consequently, there is also no action on the right--moving fermions nor on the target space spinors. Hence, this T--duality Narain Orbifold necessarily preserves all target space supersymmetries. 

In this case the parallel and perpendicular projector~\eqref{eq:InvariantSubspaceProjector} is given by
\equ{ 
\hcP^\parallel = 
\dfrac 12 
\pmtrx{
 \phantom{-}\Id_D & -\Id_D & 0 \\ 
 -\Id_D & \phantom{-}\Id_D & 0 \\ 
 \phantom{-}0 & \phantom{-}0 & \Id_{16}
}~,
\quad 
\hcP^\perp = 
\dfrac 12 
\pmtrx{
\Id_D & \Id_D & 0 \\ 
 \Id_D & \Id_D & 0 \\ 
0 & 0 & 0
}
\quad\text{and hence}\quad
V^\parallel = \pmtrx{ \phantom{-}w_\text{L} \\ -w_\text{L} \\ \phantom{-}w_{16}}~, 
}
is parameterized by integral vectors $w_\text{L}\in \Intr^D$ and $w_{16} \in \Intr^{16}$\,. The consistency condition~\eqref{eq:ProperProjectionsPartition} takes the form 
\equ{
\dfrac 14\, \Big( 
-2\, w_\text{L}^2 + w_{16}^T\, g_{16}\, w_{16} 
\Big) 
= (V^\parallel)^T \hget\, V^\parallel 
\equiv \dfrac 12\, v_{\text{L}}^2~.
}
Inserting the form~\eqref{eq:LeftNarainOrbifoldvL} of $v_{\text{L}}$ and using that $w_\text{L}^2$ is an integer, this condition may be stated as 
\equ{ 
\dfrac 14\, \Big( 
2\, w_\text{L}^2 + w_{16}^T\, g_{16}\, w_{16} 
\Big) 
\equiv \dfrac 18\, D~. 
}
Again, the combination between brackets is even. However, in this case this has stronger consequences: In order that this condition can be solved at all, the number of internal torus dimensions $D$ has to be dividable by four. Hence only for two choices of the number of internal dimensions, namely $D=4$ and $D=8$, a left--twisted T--duality Narain orbifold is possible!

According to~\eqref{eq:DimModuliSpaceNarainOrbifold}, the orbifolded Narain moduli space has dimension 
\equ{ 
\dim\big[\fM_\mathbf{P}\big] = 
\sfrac 12 D(D+16) + \sfrac 12 D(-D+16) = 16\,D~. 
} 
This suggests that the Wilson lines are not fixed by this T--duality orbifold. Hence, even performing full T--duality twists not necessarily fixes all Narain moduli. The most general invariant matrix $\cM_\text{inv}$ can be taken to be 
\equ{
\cM_\text{inv} = 
\dfrac 12 \pmtrx{ \ga & \phantom{-}\ga & \gg \\ \gd & -\gd & 0\\ \gr & \phantom{-}\gr & 2\gt}~,
}
using~\eqref{eq:InvM}. After determining it's inverse, the invariant $\Intr_2$--grading can be expressed as 
\equ{ 
\hcZ_\text{inv} = 
\pmtrx{
- \sfrac 12 \gD(1 - \gD)^{-1} & 
\Id_D + \sfrac 12 \gD(1-\gD)^{-1} & 
-(1-\gD)^{-1}\, \ga^{-1}\gg 
\\[2ex] 
\Id_D + \sfrac 12 \gD(1-\gD)^{-1} & 
- \sfrac 12 \gD(1 - \gD)^{-1} & 
(1-\gD)^{-1}\, \ga^{-1}\gg 
\\[2ex] 
\gt^{-1}\gr\, \big[1 + \sfrac 12 \gD(1-\gD)^{-1}\big] & 
-\gt^{-1}\gr\, \big[1 + \sfrac 12 \gD(1-\gD)^{-1}\big] &
(\Id_{16} - \gD')^{-1}
}~, 
}
in terms of the matrices $\gD = \ga^{-1} \gg\, \gt^{-1} \gr$ and $\gD'= \gt^{-1} \gr\, \ga^{-1} \gg$\,. (The matrix $\gd$ dropped out during the computation.) Comparing with the standard form of the $\Intr_2$--grading~\eqref{eq:Z2GradingExplicitly}, the following identifications can be made
\equ{
\ga_{16}^{-1} A = \gt^{-1} \gr~,
\quad 
A^T \ga_{16} = ( 1 - \sfrac 12 \gD)^{-1} \ga^{-1}\gg
\qand
G = \Id_D - \sfrac 12 \gD(1 - \sfrac 12 \gD)^{-1}~. 
}
The reason, that there are two expressions for the Wilson lines $A$ obtained here, reflects the fact that the procedure outlined in Subsection~\ref{sc:ConstructionOrbifoldCompatibleMetric} does not automatically enforce that $\hcH$ is symmetric. Hence this constraint has to be implemented in addition. Doing so shows that all other identifications are consistent and, in particular,
$C = \sfrac 12\, A^TA = \sfrac 12 \gD (1 - \sfrac 12 \gD)^{-1}$\,. From this it may be concluded, that the Wilson lines $A$ represent the only unfixed Narain moduli, while 
\equ{ 
G = \Id - \sfrac 12\, A^T A
\qand
B = 0~. 
}
This shows that there needs to be a bound on the Wilson lines in order to ensure that the metric $G$ remains positive definite.

The fixed points of this T--fold are determined via the equation~\eqref{eq:FixedPointCondition}
\equ{ 
\pmtrx{ 
\Id_D & \Id_D & 0 \\ 
\Id_D & \Id_D & 0 \\ 
0 & 0 & 0 
} 
Y_\text{fix} = 2 \pi\,  
\frac 12 
\pmtrx{ 
\Id_D & \Id_D & 0 \\ 
\Id_D & \Id_D & 0 \\ 
0 & 0 & 0 
} 
\pmtrx{ n \\ \tn \\ q }~. 
}
On the subspace defined by $\cP^\perp$ this can be solved as 
\equ{  \label{eq:FixedPointsLeftTwistedTfold} 
\dfrac{Y_\text{fix}}{2\gp} = \dfrac{Y_\text{fix}^\perp}{2\gp}  = 
\frac 14  
\pmtrx{ n + \tn \\ n+\tn \\ 0 } 
= 
\frac 14  
\pmtrx{ m \\ m \\ 0 }~, 
}
where $m^1,\ldots m^D=0,1,2,3$ labels $4^D$ distinct points on the doubled torus, which are not all the same as in the previous case, see~\eqref{eq:FixedPointsRightTwistedTfold}. However, in light of the residual gauge symmetry not all are physical. The orbifold compatible residual gauge transformations contain the following constant gauge parameters
\equ{ 
\gL_0 = 
\frac 12 
\pmtrx{ 
\phantom{-}\Id_D &  -\Id_D & 0 \\ 
-\Id_D & \phantom{-}\Id_D & 0 \\ 
\phantom{-}0 & \phantom{-}0 & \Id_{16} 
} 
\pmtrx{ \gm_0 \\ 0 \\ 0 } 
= 
\frac 12 \pmtrx{ \gm_0 \\ -\gm_0 \\ 0}~. 
} 
Exploiting the constant gauge transformations the upper $D$ components of the fixed point may again be set to zero by the gauge $\sfrac 12 \gm_0 = \sfrac 14 m$ and  hence a fully gauge fixed form of the fixed points reads
\equ{  \label{eq:FixedPointsLeftTwistedTfoldGaugeFixed} 
\dfrac{Y_\text{fix}^\text{g.f.}}{2\gp} =
\frac 14  
\pmtrx{ m \\  m \\ 0 }~, 
}
with $m^1,\ldots, m^D=0,1$, leading once more to $2^D$ physically distinct fixed points.

\subsection{Supersymmetric $\Intr_6$ T--fold in four dimensions}

Most T--folds discussed in the literature are variants of the T--duality orbifolds, like the ones considered above. T--folds based on other duality symmetries than $\Intr_2$ are far less known. In the following a T--fold is discussed based on the $\Intr_6$ group that acts asymmetrically on the left-- and right--movers. This example is a higher dimensional extension of the T--fold $\Intr_6$-V listed in Table 3 of~\cite{GrootNibbelink:2017usl}.

Consider the $\Intr_6$--orbifold action given by 
\equ{
\hcR = \pmtrx{
-\Id_\gd & \Id_\gd & \Id_\gd & \Id_\gd & 0 \\[1ex]
-\Id_\gd & 0 & 0 & \Id_\gd & 0 \\[1ex]
\phantom{-}\Id_\gd & 0 & 0 & 0 & 0 \\[1ex]  
-\Id_\gd & \Id_\gd & 0 & 0 & 0 \\[1ex] 
0 & 0 & 0 & 0 & \Id_{16}
}~, 
}
where $\gd$ is a positive integer for now, so that $D=2\gd$. (Later below the restriction $\gd=2$ will be considered.) This leads to an asymmetric action: 
\equ{
R_\text{R} = \pmtrx{
-\Id_\gd & 0 \\[1ex] 
0 & -\Id_\gd
}~, 
\qquad 
R_\text{L} = \pmtrx{
\sfrac 12\, \Id_\gd & -\sfrac{\sqrt 3}{2}\, \Id_\gd & 0 \\[1ex]
\sfrac{\sqrt 3}2\, \Id_\gd & \sfrac 12\, \Id_\gd & 0 \\[1ex] 
0 & 0 & \Id_{16}
}~. 
}
A $\Intr_2$--symmetry acts on the right--moving sector while a $\Intr_6$-symmetry on the left--movers. Another way to view this action is by considering the underlying $\Intr_2$ and $\Intr_3$ generators of the subgroups $\mathbf{P}_2$ and $\mathbf{P}_3$ of the point group $\mathbf{P}$: 
\equ{ 
\hcR_2 = \hcR^3 = 
\pmtrx{
-\Id_\gd & 0 & 0 & 0 & 0 \\[1ex] 
0 & -\Id_\gd & 0 & 0 & 0 \\[1ex] 
0 & 0 & -\Id_\gd & 0 & 0 \\[1ex] 
0 & 0 & 0 & -\Id_\gd & 0 \\[1ex] 
0 & 0 & 0 & 0 & \Id_{16} 
}~, 
\quad 
\hcR_3 = \hcR^2 = 
\pmtrx{
0 & 0 & -\Id_\gd & 0 & 0 \\[1ex] 
0 & 0 & -\Id_\gd & -\Id_\gd & 0 \\[1ex] 
-\Id_\gd & \phantom{-}\Id_\gd & \phantom{-}\Id_\gd & \phantom{-}\Id_\gd & 0 \\[1ex] 
0 & -\Id_\gd & -\Id_\gd & 0 & 0 \\[1ex] 
0 & 0 & 0 & 0 & \Id_{16}
}~.
}
This shows that the $\Intr_2$ action acts symmetrically and only the $\Intr_3$ is asymmetric. In fact, it only affects the left--movers: 
\equ{ 
R_{3\,\text{R}} = R_\text{R}^2 = 
\pmtrx{
\Id_\gd & 0 \\[1ex] 
0 & \Id_\gd 
}~, 
\qquad 
R_{3\,\text{L}} = R_\text{L}^2 = 
\pmtrx{
-\sfrac 12 \Id_\gd & -\sfrac{\sqrt 3}2 \Id_\gd & 0 \\[1ex] 
\phantom{-}\sfrac{\sqrt 3}2 \Id_\gd & -\sfrac 12 \Id_\gd & 0 \\[1ex] 
0 & 0 & \Id_{16}
}~. 
}

This $\Intr_6$ T--fold fixes all moduli since by \eqref{eq:DimModuliSpaceNarainOrbifold} it follows by a short computation that the dimension of the moduli space $\fM_\mathbf{P}$ is zero: 
\equ{
\text{dim}\big[ \fM_\mathbf{P} \big] = 
\dfrac 16\Big[ 
2\gd (2\gd+16) + (-2\gd)(-2\gd+16) + 2(-2\gd)(\gd+16) + 2(2\gd)(-\gd+16) 
\Big] = 0~. 
}
It is instructive to investigate which moduli are fixed by the subgroups $\mathbf{P}_2, \mathbf{P}_3 \subset \mathbf{P}$: 
\begin{subequations}
\equ{
\text{dim}\big[ \fM_{\mathbf{P}_2} \big] = 
\dfrac 12 \Big[ (2\gd)(2\gd+16) + (-2\gd)(-2\gd+16)\Big] = D^2~, 
\\[1ex]  
\text{dim}\big[ \fM_{\mathbf{P}_3} \big] = 
\dfrac 13(2\gd) \Big[ 2\gd+16 + 2(-\gd+16)\Big] = 16D~,
}
\end{subequations} 
using that $D=2\gd$. Hence, as is well--known, a symmetric $\Intr_2$ leaves all geometrical moduli, $G, B$, inert but fixes all Wilson lines $A$. The asymmetric $\Intr_3$ that only acts on the left--movers leaves the Wilson lines inert but fixes all geometrical moduli. Thus only by combining both actions all moduli, $G, B$ and $A$, together become frozen.

The metric and anti--symmetric tensor field for the geometry $\Intr_\text{6}$-V were given in Table 3 of~\cite{GrootNibbelink:2017usl}. Extending these results to the case here gives: 
\equ{
G = 
\pmtrx{
\phantom{-\sfrac 12}\Id_\gd & -\sfrac 12\Id_\gd \\[1ex]
-\sfrac 12\Id_\gd & \phantom{-\sfrac 12}\Id_\gd
}~, 
\qquad 
B = 
\pmtrx{
 0 & \sfrac 12\Id_\gd \\[1ex] 
-\sfrac 12\Id_\gd & 0 
}~, 
\qquad 
A = 0~. 
}
Inserting this in \eqref{eq:Z2GradingExplicitly} shows that the $\Intr_2$--grading reads 
\equ{
\hcZ = \dfrac 13 
\pmtrx{ 
~-\!\Id_\gd & \phantom{-}2\Id_\gd & 4\Id_\gd & \phantom{-}2\Id_\gd & 0 \\[1ex] 
-2\Id_\gd & \phantom{-2}\Id_\gd & 2\Id_\gd & \phantom{-}4\Id_\gd & 0 \\[1ex] 
\phantom{-}4\Id_\gd & -2\Id_\gd & -\!\Id_\gd & -2\Id_\gd & 0 \\[1ex]
-2\Id_\gd & \phantom{-}4\Id_\gd & 2\Id_\gd & \phantom{-2}\Id_\gd & 0 \\[1ex] 
0 & 0 & 0 & 0 & \Id_{16}
}~. 
}
By a straightforward computation it may be verified that it is a $\Intr_6$ invariant $\Intr_2$--grading: 
\equ{
\hcR^{-1}\, \hcZ\, \hcR = \hcZ
\qand
\hcZ^2 = \Id~.
}

Setting $\gd = 2$ leads to a four dimensional T--fold that preserves $\cN=1$ target space supersymmetry in the six non--compact dimensions. To see this in this formalism notice that the real and complex right--moving twist vectors can be chosen such that 
\equ{ 
v_\text{R} = \big( \sfrac 12, -\sfrac 12, \sfrac 12, -\sfrac 12 \big) 
\qand 
v_{\text{R}\Cplx} = \big( \sfrac 12, -\sfrac 12 \big)~.  
}
Hence, condition \eqref{eq:SusyCondition} can be satisfied showing that this T--fold preserves target space supersymmetry.

The condition \eqref{eq:ProperProjectionsPartition} that restricts the possible choices of $V^\parallel$ reads in this case 
\equ{ 
3\, \big(V^\parallel\big)^2  \equiv \dfrac 16~. 
} 
This is a familiar condition for four dimensional supersymmetric symmetric $\Intr_6$ orbifolds.

\appendix 
\def\theequation{\thesection.\arabic{equation}}

\section{Euclidean Pathintegrals}
\label{sc:EuclideanPathintegrals} 
\setcounter{equation}{0}

\subsection{Wick Rotation to an Euclidean Worldsheet}

The coordinate transformation
\equ{ 
\gs_1 \mapsto \sqrt 2\, z_1 
\qand 
i\,\gs_0 \mapsto \sqrt 2\, z_2 
}
defines a Wick--rotation from the Minkowskian to the Euclidean description. In terms of complex coordinates 
\equ{ 
z = z_1+i\, z_2
\qand
\bz = z_1-i\, z_2
}
this Wick--rotation reads
\equ{
\gs \mapsto z
\qand 
\bgs \mapsto \bz~. 
}
Consequently, the Minkowskian derivatives can be mapped to derivatives of these real Euclidean coordinates
\equ{
\der_0 \mapsto \dfrac{i}{\sqrt{2}} \der_2 = -\dfrac 1{\sqrt 2}(\der-\bder) 
\qand
\der_1 \mapsto \dfrac 1{\sqrt 2} \der_1 = \dfrac 1{\sqrt 2}\, (\der+\bder) 
}
The Euclidean action $S_\text{E} = -i\,S$ is obtained from the Minkowskian action $S$ after applying these substitutions.

\subsection{Euclidean Worldsheet Torus}

At the one-loop level the worldsheet is a torus $T^2_\text{WS}$ defined by the periodicities
\equ{ 
z \sim z + 1
\qand
z \sim z + \gt 
}
and, consequently, the area of the worldsheet torus equals 
\equ{
\text{Vol}(T^2_\text{WS}) = \int \d^2 z = 2 \gt_2~. 
}
By expressing this complex coordinate $z$ and its conjugate $\bz$ as 
\equ{  \label{eq:StraightWSTcoord}
z = x + \gt\, y
\qand
\bz = x + \bgt\, y~, 
}
in terms of the real coordinates $(x,y)$, the periodicities can be stated as 
\equ{
x \sim x+1
\qand
y \sim y+1~.
}

\subsection{Mode Functions on an Euclidean Worldsheet Torus}
\label{sc:ModeFunctions}

The functions 
\equ{ \label{eq:SimpleMode}
\gf_1(z) = \dfrac{ \gt\, \bz- \bgt\,z}{\gt - \bgt}~; 
\quad
\gf_\gt(z) = \dfrac{\bz- z}{\gt - \bgt}~, 
}
are quasi--periodic in single directions on the wordsheet torus
\equ{
\gf_1(z+1) = \gf_1(z) +1~,~
\gf_1(z-\gt) = \gf_1(z)~;~~
\gf_\gt(z+1) = \gf_\gt(z)~,~
\gf_\gt(z-\gt) = \gf_\gt(z)+1~. 
}
Note that 
\begin{subequations} 
\equa{
(\der+\bder) \gf_1(z) = 1~, 
\qquad 
(\der-\bder) \gf_1(z) = i\, \dfrac{\gt_1}{\gt_2}~, 
\\[1ex] 
(\der+\bder) \gf_\gt(z) = 0~, 
\qquad 
(\der-\bder) \gf_\gt(z) = i\, \dfrac{1}{\gt_2}~. 
}
\end{subequations} 
Hence, the functions 
\equ{ \label{eq:LinearMode}
\gf\brkt{r}{r'}(z) = r\, \gf_1(z) + r'\,\gf_\gt(z)
\qand
\gf\brkt{r}{r'}(x,y) = r\, x - r'\,y~,
}
where $r, r' \in \Ratl$, possess the following properties on the wordsheet torus
\equ{
\gf\brkt{r}{r'}(z+1) = \gf\brkt{r}{r'}(z) + r
\qand
\gf\brkt{r}{r'}(z-\gt) = \gf\brkt{r}{r'}(z) + r'~,  
}
so that the mode functions 
\equ{ \label{eq:QuasiPeriodicMode} 
\gF\brkt{r}{r'}(z) = e^{2\pi i \gf\brkt{r}{r'}(z)} 
}
are quasi--periodic 
\equ{ 
\gF\brkt{r}{r'}(z+1) = e^{2\pi i\, r}\, \gF\brkt{r}{r'}(z)
\qand
\gF\brkt{r}{r'}(z+\gt) = e^{-2\pi i\, r'}\, \gF\brkt{r}{r'}(z)~. 
}
These mode functions are orthogonal in the sense that 
\equ{ 
\int \d^2z\, \overline{ \gF\brkt{r}{r'}(z) }\, \gF\brkt{s}{s'}(z)
= 2\gt_2\, \gd^{rs}\, \gd_{r's'}~.  
} 

\subsection{Left--Moving Fermionic Partition Functions}
\label{sc:FermionicPartitionFunctions} 

Consider a $d$--dimensional vector of complex left--moving fermions $\gvf_\Cplx=(\gvf_1,\ldots,\gvf_d)$ with quasi--periodic boundary conditions
\equ{
\gvf_\Cplx(z+1) = e^{2\pi i\, (\Bga + \sfrac s2\Id_d)}\, \gvf_\Cplx(z)
\qand
\gvf_\Cplx(z-\gt) = e^{2\pi i\, (\Bga'+ \sfrac{s'}2\Id_d)}\, \gvf_\Cplx(z)~, 
}
labelled by two real $d$--dimensional vectors $\ga,\ga'$ and spin structures $s,s'=0,1$\,. In the exponentials $\Bga$ denotes the diagonal $d\times d$--matrix,  
\equ{ \label{eq:MatrixTwistVector}
\Bga = \pmtrx{
\ga_1 & 0 & \cdots & 0 \\
0 & \ddots & \ddots & \vdots \\
\vdots & \ddots & \ddots & 0 \\
0 & \cdots &  0 & \ga_d 
}~, 
}
where diagonal entries are the components of the vector $\ga=(\ga_1,\ldots,\ga_d)$\,. These boundary conditions can be solved by the complex mode expansion 
\equ{
\gvf_\Cplx(z) = \sum_{r,r'} \gF\brkt{\Bga\phantom{'}+(r\phantom{'}+\sfrac{s\phantom{'}}2)\Id_d}{\Bga'+(r'+\sfrac{s'}2)\Id_d}(z)\, 
\gvf\brkt{r}{r'}~, 
}
exploiting the mode functions~\eqref{eq:QuasiPeriodicMode}.

Using this mode expansion the Euclidean pathintegral 
\equ{ 
Z_{\Cplx^d\,\text{Ferm}}\brkt{\ga\phantom{'},s\phantom{'}}{\ga',s'} 
= \cD \bgvf_\Cplx \cD \gvf_\Cplx\, 
e^{S_{\Cplx\,\text{Ferm}}}~,   
\quad
S_{\Cplx\,\text{Ferm}} = \int \dfrac{\d^2z}{2\pi}\, 
\bgvf_\Cplx\, \bder \gvf_\Cplx~, 
}
can be expressed as infinite product 
\equ{
Z_{\Cplx^d\,\text{Ferm}}\brkt{\ga\phantom{'},s\phantom{'}}{\ga',s'}  
\sim 
\prod_{r,r'} \det\big[ (\Bga+(r+\sfrac{s}2)\Id_d)\gt + \Bga'+(r'+\sfrac{s'}2)\Id_d\big]~, 
}
which can be regularized up to a constant factor to the genus--$d$ quotient of a theta--function~\eqref{eq:Theta} and a Dedekind function~\eqref{eq:Dedekind}: 
\equ{ \label{eq:PartitionCFerm}
Z_{\Cplx^d\,\text{Ferm}}\brkt{\ga\phantom{'},s\phantom{'}}{\ga',s'}   
\sim 
 \frac{\gth_d\brkt{\sfrac12 e_d - \sfrac {s}{2} e_d- \ga}{\sfrac 12 e_d -\sfrac {s'}{2} e_d-\ga'}(\gt)}{\get^d(\gt)}~, 
}
where $e_d^T = (1,\ldots, 1)$ is the $d$--dimensional vector with only ones.

Performing the sum over the spin structures $s,s'$ with appropriate phases, the partition functions
\equ{ \label{eq:LMpartitionFun} 
Z_{\Cplx^d\,\text{Ferm}}\brkt{\ga}{\ga'}(\gt) = 
e^{-\pi i\, \ga^T(\ga'-e_d)}\, \frac 1{2} \sum_{s,s'=0}^{1}\, e^{-\pi i\, \sfrac 1{4}e_d^Te_d\, s's}\, 
e^{-2\pi i\, \sfrac{s'}2 e_d^T \ga} \, 
 \frac{\gth_d\brkt{\sfrac12 e_d - \sfrac {s}{2} e_d- \ga}{\sfrac 12 e_d -\sfrac {s'}{2} e_d-\ga'}(\gt)}{\get^d(\gt)}~, 
}
are obtained, that are modular covariant in the sense that 
\begin{subequations}
\equa{ \label{eq:ModCovTransPartition}
Z_{\Cplx^d\,\text{Ferm}}\brkt{\ga}{\ga'}\big(\gt+1\big) &= e^{2\pi i\, \sfrac{d}{12}}\, Z_{\Cplx^d\,\text{Ferm}}\brkt{\ga}{\ga'+\ga}\big(\gt\big)~, 
\\[1ex] 
Z_{\Cplx^d\,\text{Ferm}}\brkt{\ga}{\ga'}\big(\sfrac{-1}\gt\big) &= 
e^{-2\pi i\, \sfrac d4}\, 
Z_{\Cplx^d\,\text{Ferm}}\brkt{\ga'}{-\ga}\big(\gt\big)~,
}
\end{subequations} 
assuming that $d/4 \in \Natr$. 

In the main text we are not concerned with $d$ complex fermions but with $d$ real fermions. To be able to deal with real fermions in much as the same fashion as complex fermions, consider the complexification of the real fermions. The partition function for the real fermions can then be defined as the square root of the complexified fermions: 
\equ{
Z_{\Real^d\,\text{Ferm}}\brkt{\ga\phantom{'},s\phantom{'}}{\ga',s'} = 
\Big[Z_{\Cplx^d\,\text{Ferm}}\brkt{\ga\phantom{'},s\phantom{'}}{\ga',s'}\Big]^{1/2}  
\sim 
 \frac{\Big[\gth_d\brkt{\sfrac12 e_d - \sfrac {s}{2} e_d- \ga}{\sfrac 12 e_d -\sfrac {s'}{2} e_d-\ga'}(\gt)\Big]^{1/2}}{\get^{d/2}(\gt)}~, 
}
in analogy to~\eqref{eq:PartitionCFerm}. The complex branch here is chosen such that if the complex partition function transforms with a phase $e^{i\,\go}$ then the real partition function picks up the phase $e^{i\, \go/2}$\,. 
Furthermore, for all the inner products $\ga^T\gb$ of two $d$--dimensional vectors in the additional phase factors in~\eqref{eq:LMpartitionFun} acquire an additional factor $1/2$. Hence, \eqref{eq:LMpartitionFun} for $d$ real fermions reads: 
\equ{ \label{eq:LMpartitionFunReal} 
Z_{\Real^d\,\text{Ferm}}\brkt{\ga}{\ga'}(\gt) = 
e^{-\pi i\, \sfrac 12\ga^T(\ga'-e_d)}\, \frac 1{2} \sum_{s,s'=0}^{1}\, e^{-\pi i\, \sfrac 1{8}e_d^Te_d\, s's}\, 
e^{-2\pi i\, \sfrac{s'}4 e_d^T \ga} \, 
 \frac{\Big[\gth_d\brkt{\sfrac12 e_d - \sfrac {s}{2} e_d- \ga}{\sfrac 12 e_d -\sfrac {s'}{2} e_d-\ga'}(\gt)\Big]^{1/2}}{\get^{d/2}(\gt)}~, 
}
which is modular covariant in the sense that 
\begin{subequations}
\equa{ \label{eq:ModCovTransPartitionReal}
Z_{\Real^d\,\text{Ferm}}\brkt{\ga}{\ga'}\big(\gt+1\big) &= e^{2\pi i\, \sfrac{d}{24}}\, \mathcal{Z}_{\Real^d\,\text{Ferm}}\brkt{\ga}{\ga'+\ga}\big(\gt\big)~, 
\\[1ex] 
Z_{\Real^d\,\text{Ferm}}\brkt{\ga}{\ga'}\big(\sfrac{-1}\gt\big) &= 
e^{-2\pi i\, \sfrac d8}\, 
Z_{\Real^d\,\text{Ferm}}\brkt{\ga'}{-\ga}\big(\gt\big)~,
}
\end{subequations} 
assuming that $d/8 \in \Natr$.

\subsection{Left--Moving Chiral Bosonic Partition Functions}

Consider a $d$--dimensional vector of complex left--moving chiral bosons $y_\Cplx=(\gvf_1,\ldots,\gvf_d)$ with quasi-periodic boundary conditions
\equ{
y_\Cplx(z+1) = e^{2\pi i\, \Bga}\, y_\Cplx(z)
\qand
y_\Cplx(z-\gt) = e^{2\pi i\, \Bga'}\, y_\Cplx(z)~, 
}
using the same notation as introduce above. These boundary conditions can be solved by the complex mode expansion 
\equ{
\gvf_\Cplx(z) = \sum_{r,r'} \gF\brkt{\Bga\phantom{'}+r\phantom{'}\,\Id_d}{\Bga'+r'\,\Id_d}(z)\, 
y\brkt{r}{r'}~, 
}
exploiting again the mode functions~\eqref{eq:QuasiPeriodicMode}. 

Assuming that there no $i$ such that both $\ga_i$ and $\ga_i'$ are integral at the same time, the Euclidean pathintegral 
\equ{ 
Z_{\Cplx^d\,\text{Ch.Bos}}\brkt{\ga\phantom{'}}{\ga'} 
= \cD \byy_\Cplx \cD y_\Cplx\, 
e^{S_{\Cplx\,\text{Ch.Bos}}}~,   
\quad
S_{\Cplx\,\text{Ch.Bos}} = \int \dfrac{\d^2z}{2\pi}\, 
 \dfrac{-1}{\sqrt 2} \,\der_1 \byy_\Cplx\, \bder y_\Cplx~, 
}
can be expressed as infinite product 
\equ{ \label{eq:ChBosonPartition} 
Z_{\Cplx^d\,\text{Ch.Bos}}\brkt{\ga\phantom{'}}{\ga'}  
\sim 
\prod_{r,r'} \dfrac{1}{\det\big[ r (\Bga+r\,\Id_d)\gt + \Bga'+r'\,\Id_d\big]}
\sim 
 \frac{\get^d(\gt)}{\gth_d\brkt{\sfrac12 e_d - \ga}{\sfrac 12 e_d-\ga'}(\gt)}~, 
}
where the infinite factor $\prod_{r,r'} r$ has been dropped.

Including appropriate phases, the partition functions
\equ{ \label{eq:BosLMpartitionFun} 
Z_{\Cplx^d\,\text{Ch.Bos}}\brkt{\ga}{\ga'}(\gt) = 
e^{\pi i\, \ga^T(\ga'-e_d)}\, 
 \frac{\get^d(\gt)}{\gth_d\brkt{\sfrac12 e_d - \ga}{\sfrac 12 e_d -\ga'}(\gt)}~, 
}
are obtained, that are modular covariant in the sense that 
\begin{subequations}
\equa{ \label{eq:ModCovTransBosPartition}
Z_{\Cplx^d\,\text{Ch.Bos}}\brkt{\ga}{\ga'}\big(\gt+1\big) &= e^{-2\pi i\, \sfrac{d}{12}}\, Z_{\Cplx^d\,\text{Ch.Bos}}\brkt{\ga}{\ga'+\ga}\big(\gt\big)~, 
\\[1ex] 
Z_{\Cplx^d\,\text{Ch.Bos}}\brkt{\ga}{\ga'}\big(\sfrac{-1}\gt\big) &= 
e^{2\pi i\, \sfrac d4}\, 
\mathcal{Z}_{\Cplx^d\,\text{Ch.Bos}}\brkt{\ga'}{-\ga}\big(\gt\big)~.
}
\end{subequations} 

For real chiral bosons we can proceed as for the real fermions, hence the partition function for $d$ real chiral bosons reads: 
\equ{ \label{eq:LMBosPartitionFunReal} 
Z_{\Real^d\,\text{Ch.Bos}}\brkt{\ga}{\ga'}(\gt) = 
e^{-\pi i\, \sfrac 12\ga^T(\ga'-e_d)}\, 
 \frac{\get^{d/2}(\gt)}{\Big[\gth_d\brkt{\sfrac12 e_d - \ga}{\sfrac 12 e_d-\ga'}(\gt)\Big]^{1/2}}~, 
}
which is modular covariant in the sense that 
\begin{subequations}
\equa{ \label{eq:ModCovTransBosPartitionReal}
Z_{\Real^d\,\text{Ch.Bos}}\brkt{\ga}{\ga'}\big(\gt+1\big) &= e^{-2\pi i\, \sfrac{d}{24}}\, Z_{\Real^d\,\text{Ch.Bos}}\brkt{\ga}{\ga'+\ga}\big(\gt\big)~, 
\\[1ex] 
Z_{\Real^d\,\text{Ch.Bos}}\brkt{\ga}{\ga'}\big(\sfrac{-1}\gt\big) &= 
e^{+2\pi i\, \sfrac d8}\, 
Z_{\Real^d\,\text{Ch.Bos}}\brkt{\ga'}{-\ga}\big(\gt\big)~.
}
\end{subequations} 

\section{Modular Functions and Transformations}
\label{sc:ModFunTrans} 
\setcounter{equation}{0}

The complex Teichm\"uller parameter $\gt$ subject to modular transformations generated by 
\equ{ \label{eq:ModTrans}
\gt \mapsto \gt+1
\qand 
\gt \mapsto \dfrac {-1}{\gt}~,  
}
so that  
\equ{ \label{eq:ModTransReIm}
\gt_1 \mapsto \gt_1~, 
\quad 
\gt_2 \mapsto \gt_2
\qand
\gt_1 \mapsto \dfrac {-1}{|\gt|^2}\, \gt_1~, 
\quad 
\gt_2 \mapsto \dfrac {1}{|\gt|^2}\, \gt_2~. 
}

The Dedekind function is given by 
\equ{ \label{eq:Dedekind} 
\get(\gt) = q^{\frac 1{24}} \prod_{n\geq 1} \big( 1 - q^n\big)~, 
\quad
\text{with}
\quad 
q = e^{2\pi i\, \gt}~. 
}
The Dedekind function transforms under the generators of the modular transformations as 
\equ{ \label{eq:ModTransDedekind} 
\get\big(\gt+1\big) = e^{2\pi i\, \frac 1{24}}\,\get\big(\gt\big)
\qand
\get\Big( \frac {-1}{\gt} \Big) = 
(-i\gt)^{1/2}\, \get\big(\gt\big)~. 
}
The Dedekind function may be related to the infinite product
\equ{ \label{eq:ProductDedekind}
\left.\prod_{r,r'}\right.^\prime \dfrac{1}
 {r\, \gt + r'}
= \dfrac 1{
\get(\gt)}~. 
}

Another important modular functions are the genus--$d$ theta functions 
\equ{ \label{eq:Theta}
\gth_d\brkt{\ga}{\ga'}\big(z\big|\gt\big) = 
\sum_{n\in\Intr^d} q^{\frac 12\, (n-\ga)^2}\, 
e^{2\pi i\, (z-\ga')^T(n-\ga)}~, 
}
with the $d$--dimensional vector valued characteristics $\ga$ and $\ga'$ and argument $z$\,. (Often $z=0$ and then this argument is dropped.) These characteristics have the following quasi--periodicities 
\equ{
\gth_d\brkt{\ga+n}{\ga'}\big(z\big|\gt\big) = 
\gth_d\brkt{\ga}{\ga'}\big(z\big|\gt\big)
\qand
\gth_d\brkt{\ga}{\ga'+n}\big(z\big|\gt\big) = 
e^{2\pi i\, n^T\ga}\, \gth_d\brkt{\ga}{\ga'}\big(z\big|\gt\big)~, 
}
for $n \in \Intr^d$. Their modular transformations read
\begin{subequations}
\equa{ 
\gth_d\brkt{\ga}{\ga'}\big(z\big|\gt+1\big) &= 
e^{- \gp i\, \ga^T(\ga+e_d)}\, 
\gth_d\brkt{\ga}{\ga'+\ga+\frac 12\, e}\big(z\big|\gt\big)~, 
\\[1ex] 
\gth_d\brkt{\ga}{\ga'}\big(\dfrac{z}{\ga}\big|\frac{-1}{\gt}\big) &= 
(-i\gt)^{d/2}\, e^{2\pi i\big( \frac{z^2}{2\gt} +\ga^T\ga'\big)}
\gth_d\brkt{\ga'}{-\ga}\big(z\big|\gt\big)~, 
}
\end{subequations} 
where $e_d$ is defined under \eqref{eq:PartitionCFerm}.

\section{Narain Geometry}
\label{sc:Narain Geometry}
\setcounter{equation}{0}

\subsection{Properties of the T--duality Group}

Narain tori and orbifolds concern themselves with lattice vectors $N\in\Intr^{2D+16}$. A general lattice transformation is given by 
\equ{
N \mapsto \hcM\, N~, 
\quad
\hM \in \text{GL}(2D+16;\Intr)~. 
}
In order that the matrix $\hcM$ is invertible over the integers, it's determinant $\det[\hcM] = \pm 1$. 

The metric~\eqref{eq:ODD16Metric} is defined in terms of the 16-dimensional Cartan metric and vielbein  
\equ{ \label{eq:E88metric}
g_{16}=\ga_{16}^T\, \ga_{16}~, 
\quad\text{with}\quad 
\ga_\text{16} = \pmtrx{ \ga_{8} & 0 \\[1ex] 0 & \ga_{8} }
}
of the even, self--dual $\E{8}\times\E{8}$--root lattice are introduced in terms of the matrix of eight simple root vectors of $\E{8}$ given by 
\equ{ \label{eq:E8roots} 
\ga_8 = 
\pmtrx{ 
~1 & 0 & 0 & 0   & 0 & 0 & 0 & \sm \sfrac 12 
\\ 
\sm 1 & ~1 & 0 & 0   & 0 & 0 & 0 & \sm \sfrac 12 
\\
0 &  \sm 1 & ~1 & 0  & 0 & 0 & 0 & \sm \sfrac 12 
\\ 
0 & 0 & \sm 1 & ~1   & 0 & 0 & 0 & \sm \sfrac 12 
\\ 
0 & 0 & 0 & \sm 1  & ~1 & 0 & 0 & \sm \sfrac 12 
\\ 
0 & 0 & 0 & 0 & \sm 1 & ~1 & 1 & \sm \sfrac 12 
\\ 
0 & 0 & 0 & 0 & 0 & \sm 1 & 1 & \sm \sfrac 12 
\\
0 & 0 & 0 & 0 & 0 & 0 & 0 & \sm \sfrac 12 
}~. 
}
Note that $\det[\ga_8]=-1$\,, hence $\det[g_{16}] = \det[\ga_{16}] = 1$ and $\det[\hget] = (-1)^D$\,. Consequently, all the matrices $\ga_8, \ga_{16}, g_{16}$ and $\hget$ are invertible over the integers.

The subgroup $\text{O}_{\hget}(D,D+16;\Real) \subset \text{GL}(2D+16;\Real)$ is defined by requiring that its matrices leave the metric $\hget$ invariant, i.e.\ $\hcM \in \text{O}_{\hget}(D,D+16;\Real)$ when
\equ{
\hcM^T \hget\, \hcM = \hget
}
for $\hcM\in\text{GL}(2D+16;\Real)$\,. The integral group $\text{O}_{\hget}(D,D+16;\Intr)$ is defined in the same fashion, but then $\hcM\in\text{GL}(2D+16;\Intr)$\,. This group is often referred to as the T--duality group. When $\hcM\in\text{GL}(2D+16;\Intr)$ but not in $\text{O}_{\hget}(D,D+16;\Intr)$ the metric $\hget$ takes a different form.

\subsection{Parameterizing Narain Moduli}

\subsubsection*{Generalized Narain Metric and $\Intr_2$--Grading}

The generalized metric $\hcH$, defined in~\eqref{eq:GeneralizedNarainMetric}, contains the following Narain moduli: a $D$--dimensional metric $G = G^T \in \text{GL}(D; \Real)$, an anti--symmetric tensor $B=-B^T \in \text{M}_{D\times D}(\Real)$ and a Wilson line matrix $A \in \text{M}_{16\times D}(\Real)$. The latter can be thought of as $D$ Wilson lines $A_i$ on the $D$ dimensional torus directions combined into a single matrix $A$. The generalized metric transforms covariantly as 
\equ{
\hcH \mapsto 
\hcM^T \hcH\, \hcM
}
under $\hcM\in\text{O}_{\hget}(D,D+16;\Real)$\,. Related to the generalized metric is the $\Intr_2$--grading
\equ{\label{eq:Z2GradingExplicitly}
\hcZ = \hget^{-1} \hcH =  
\pmtrx{
-G^{-1} C^T & G^{-1} & - G^{-1} A^T \ga_\text{16}
\\[1ex] 
G + A^T A + C G^{-1} C^T & -C G^{-1} & (\Id_D + C G^{-1}) A^T \ga_\text{16}
\\[1ex] 
\ga_\text{16}^{-1} A (\Id_D+ G^{-1} C^T) & -\ga_\text{16}^{-1} A G^{-1} & \ga_\text{16}^{-1} \big(\Id_{16}+ A G^{-1} A^T \big) \ga_\text{16}
}~.
}

\subsubsection*{Generalized Narain Vielbein}

A generalized vielbein
\equ{ \label{eq:NarainVielbein} 
\hcE = \pmtrx{
e & 0 & 0 
\\[1ex]
-e^{-T} C^T & e^{-T} & -e^{-T} A^T \ga_{16}
\\[1ex] 
\ga_{16}^{-1} A & 0 & \Id_{16} 
}
}
is itself an element of $\text{O}_{\hget}(D,D+16;\Real)$\,, since 
\equ{
\hcE^T\, \hget\, \hcE = \hget~. 
}
In particular, it's inverse reads: 
\equ{ \label{eq:InvNarainVielbein} 
\hcE^{-1} = \hget^{-1} \hcE^T\, \hget 
= \pmtrx{
e^{-1} & 0 & 0 \\[1ex] 
- C e^{-1} & e^T & A^T\ga_{16} \\[1ex] 
-\ga_{16}^{-1} A e^{-1} & 0 & \Id_{16} 
}~. 
}
The matrix $\cK$ relates the metric $\hget$ to the $(D,D+16)$--dimensional Minkowski metric $\get$ 
\equ{ \label{eq:MinkowskiDD16Metric} 
\cK^{-T}\, \hget\, \cK^{-1} = \get = 
\pmtrx{-\Id_D & 0 &0 \\[1ex] 0 & \Id_{D} & 0 \\[1ex] 0 & 0 & \Id_{16} }~,
\quad \text{where} \quad 
\cK = \pmtrx{\frac 1{\sqrt 2}\,\Id_D & -\frac 1{\sqrt 2}\,\Id_D & 0 \\[1ex] \frac 1{\sqrt 2}\,\Id_D & \phantom{-}\frac 1{\sqrt 2}\,\Id_D & 0 \\[1ex] 0 & 0 & \ga_{16}}~. 
}
Hence 
\equ{ \label{eq:PropertyGeneralizedVielbein} 
\cE^T \get\, \cE = \hget~, 
\quad\text{where}\quad
\cE = \cK\, \hcE~. 
}
The generalized vielbein can be thought of as the square root of the generalized metric in the sense that 
\equ{
\hcH = \cE^T\, \cE = \hcE^T \hat\Id\, \hcE
\quad\text{where}\quad
\hat\Id = \cK^T\cK = 
\pmtrx{ \Id_D & 0 & 0 \\ 0 & \Id_D & 0 \\ 0 & 0 & g_{16}}~.  
}

\subsubsection*{Narain Vielbein Decomposition}

The generalized vielbein~\eqref{eq:NarainVielbein} can be decomposed as follows 
\equ{\label{eq:NarainVielbeinDecomposition}
\hcE(e,B,A) = \hcM_{e}(e)\, \hcM_b(B)\, \hcM_a(A)~, 
}
where the following $\text{O}_{\hget}(D,D+16;\Real)$--matrices are introduced
\equ{ \label{eq:GeometricModuliMatrices} 
\hcM_{e}(e) = 
\pmtrx{ 
e & 0 & 0 \\[1ex]
0 & e^{-T} & 0 \\[1ex]
0 & 0 & \Id_{16} 
}, ~
\hcM_{b}(b) = 
\pmtrx{ 
\Id_D & 0 & 0 \\[1ex]
b & \Id_D & 0 \\[1ex]
0 & 0 & \Id_{16} \\[1ex]
},~
\hcM_{a}(a) = 
\pmtrx{
\Id_D & 0 & 0 \\[1ex]
-\frac{1}{2} a^T  a  & \Id_D & -a^T \ga_\text{16} \\[1ex]
\ga_\text{16}^{-1} a & 0 & \Id_{16} 
}, 
}
where $e \in \text{GL}(D;\Real)$ is an invertible $D\times D$--matrix, $b = - b^T \in \text{M}_{D\times D}(\Real)$ an anti--symmetric $D\times D$--matrix and $a \in \text{M}_{D\times D}(\Real)$ a $16\times D$--matrix. Here the notation is such that the subscripts identify the type of transformations and the arguments the parameters of these transformations. The product of two Narain vielbeins read 
\begin{subequations} \label{eq:ProductNarainVielbeins}
\equ{
\hcE(E,B,A)\, \hcE(e,b,a) = \hcE(E', B', A')~, 
}
where 
\equ{ 
E' = E\, e~, 
\quad 
B' = e^T B e + b + \sfrac 12 e^T A^T a- \sfrac 12 a^T A e
\qand
A' = A e + a~, 
} 
\end{subequations} 
using the multiplication rules obtained in~\cite{GrootNibbelink:2017usl}.

The T--duality operator in all $D$ directions can be introduced as 
\equ{ \label{eq:FullTDualityOperator} 
\hI 
=
\cK^{-1} \get\, \cK = 
\pmtrx{
0 &  \Id_D & 0 \\[1ex]
\Id_D & 0 & 0 \\[1ex]
0 & 0 & \Id_{16} \\[1ex]
}~, 
}
where $\cK$ is introduced in~\eqref{eq:MinkowskiDD16Metric}. 
Clearly, it squares to the identity. It can be used to define the following non--geometric $\text{O}_{\hget}(D,D+16;\Real)$--transformations 
\equ{  \label{eq:NonGeometricModuliMatrices} 
\hcM_\gb(\gb) = \hI\, \hcM_b(\gb)\, \hI 
= \pmtrx{
\Id_D &  \beta & 0 \\[1ex]
0 & \Id_D & 0 \\[1ex]
0 & 0 & \Id_{16} \\[1ex]
},~ 
\hcM_\ga(\ga) = \hI\, \hcM_a(\ga)\, \hI 
= \pmtrx{
\Id_D & -\frac 12  \ga^T \ga  & -  \ga^T \ga _\text{16}
\\[1ex] 
0 & \Id_D & 0 
\\[1ex] 
0 & \ga_\text{16}^{-1} \ga & \Id_{16}
}, 
}
where $\beta = -\beta^T \in \text{M}_{D\times D}(\Real)$ and $\ga \in \text{M}_{16\times D}(\Real)$\,.

\subsection{Generalized Metric Independence}
\label{sc:GenMetricIndependence} 

Consider two invertible $(2D+16)\times (2D+16)$--matrices $\cA$ and $\cB$ that mutually commute and commute with the $\Intr_2$--grading $\hcZ=\hget^{-1}\hcH$. They appear in the determinant 
\equ{
\det\big[ \hget\,\cA + \hcH\, \cB\big] 
= 
\det\big[\hget \big( \Id + \hget^{-1}\hcH\, \cB \cA^{-1}\big) A\big] 
}
which often arises in Gaussian pathintegrals over the Narain coordinate fields $Y$. This can be rewritten as
\equ{ \label{eq:DetExpression} 
\det\big[ \hget\,\cA + \hcH\, \cB\big] = \det \hget\, \det \cA\, 
\det\big[ \Id + \cC\, \hcZ\big]~, 
}
where $\cC = \cB\cA^{-1}$ commutes with $\hcZ=\hget^{-1}\hcH$. By taking the log of the final determinant factor and using the Taylor expansion of the ln--function this expression takes the form
\equ{ 
\ln\det\big[ \Id + \cC \hcZ\big] = \tr\ln\big[ \Id + \cC \hcZ\big] = 
\tr\Big[
\sum_{m\geq 1} \dfrac {-1}{2m}\, \cC^{2m} 
+
\sum_{m\geq 0} \dfrac {1}{2m+1}\, \cC^{2m+1}\,\hcZ
\Big]
}
since $\hcZ$ squares to unity. Using that the left-- and right--moving projectors~\eqref{eq:SplittingNarainOperators} satisfy 
\equ{
\Id = \hcP_\text{L} + \hcP_\text{R}
\qand
\hcZ = \hcP_\text{L} - \hcP_\text{R}~, 
}
this can be split in left-- and right--moving contributions
\equ{
\ln\det\big[ \Id + \cC \hcZ\big] =
\tr_\text{L}\big[ \ln\big(\Id + \cC)\big] + \tr_\text{R}\big[ \ln \big(\Id + \cC)\big]~,
}
where the projected traces $\tr_\text{L/R} [\cX] = \tr\big[ \cX \hcP_\text{L/R}\big]$ were introduced. Hence, the initial expression~\eqref{eq:DetExpression} can be written as 
\equ{ \label{eq:DetRelationNarain}
\det\big[ \hget\,\cA + \hcH\, \cB\big] 
=  \det \hget\, \det{}_\text{L}[\cA+\cB]\, \det{}_\text{R}[\cA-\cB]~, 
}
in terms of determinants $\det{}_\text{L/R}$ over the $D+16$-- and $D$--dimensional subspaces defined by the projections $\hcP_\text{L/R}$. This result shows that this determinant is independent of the (fixed or free) moduli contained in the generalized metric $\hcH$. These determinants may be evaluated in the lattice basis but also in the basis where the orbifold twists are (block) diagonal, e.g.~\eqref{eq:BlockDD16Matrices}.

Taking 
\equ{
\cA = \gt_1\, \mathbf{a} + \mathbf{a}'
\qand 
\cB = i \gt_2\, \mathbf{a}~, 
}
for two commuting matrix $\mathbf{a}$ and $\mathbf{a}'$, this leads to the identity 
\equ{ \label{eq:DetRelationNarainII}
\det\big[ \hget\,(\mathbf{a}\,\gt_1+\mathbf{a}') + \hcH\, i\gt_2\, \mathbf{a} \big] 
=  \det \hget\, \det{}_\text{L}[\mathbf{a}\,\gt + \mathbf{a}']\, \det{}_\text{R}[\mathbf{a}\,\bgt+\mathbf{a}']~. 
}
\bibliographystyle{paper}
{\small
\bibliography{paper}
}

\end{document}